\documentclass[amsmath,amssymb,aps,pra,reprint,superscriptaddress,longbibliography]{revtex4-2}

\usepackage[svgnames,psnames]{xcolor}
\usepackage{tabularx}
\usepackage{floatrow}
\usepackage{graphicx}
\usepackage[label font=large]{subfig}
\usepackage{url}
\usepackage{braket}
\usepackage{amsmath,amssymb}
\usepackage{graphicx}% Include figure files
\graphicspath{{.}{figs/}{../figs/}}
\usepackage{afterpage}
\usepackage{dcolumn}% Align table columns on decimal point
\usepackage{bm}% bold math
\usepackage{tikz}
\usetikzlibrary{positioning,intersections,angles,quotes}
\usepackage{comment}
\usepackage{physics}
\usepackage{mathtools}
\usepackage{chemfig}
\usepackage[version=4]{mhchem}
\usepackage[colorlinks,citecolor=DarkGreen,linkcolor=FireBrick,linktocpage,unicode,hypertexnames=false]{hyperref}
\newtheorem{theorem}{Theorem}
\usepackage{ulem}
\usepackage{quantikz}
\usepackage{adjustbox}
\usepackage{booktabs}

\usepackage{algorithm}
\usepackage{algorithmic}

\makeatletter
\def\@opargbegintheorem#1#2#3{\trivlist
   \item[]{\bfseries #1\ #2\ (#3)} \itshape}
\makeatother

\DeclareFloatSeparators{mysep}{\hskip-30em}

\usepackage{enumerate}

\newcommand{\qunasys}{QunaSys, 1-13-7 Hakusan, Bunkyo-ku, Tokyo 113-0001, Japan}
\newcommand{\icepp}{International Center for Elementary Particle Physics, The University of Tokyo, 7-3-1 Hongo, Bunkyo-ku, Tokyo 113-0033, Japan}

\begin{document}

\title{Quantum Power Iteration Unified Using Generalized Quantum Signal Processing}
\author{Viktor Khinevich}%
\email{victorkh711@gmail.com}
\affiliation{%
  Graduate School of Engineering Science, The University of Osaka, 1-3 Machikaneyama, Toyonaka, Osaka 560-8531, Japan
}%
\affiliation{%
  Center for Quantum Information and Quantum Biology,
  The University of Osaka, 1-2 Machikaneyama, Toyonaka 560-8531, Japan
}%

\author{Yasunori Lee}
\affiliation{\qunasys}

\author{Nobuyuki Yoshioka}
\affiliation{\icepp}

\author{Wataru Mizukami}%
\email{mizukami.wataru.qiqb@osaka-u.ac.jp}
\affiliation{%
  Graduate School of Engineering Science, The University of Osaka, 1-3 Machikaneyama, Toyonaka, Osaka 560-8531, Japan
}%
\affiliation{%
  Center for Quantum Information and Quantum Biology,
  The University of Osaka, 1-2 Machikaneyama, Toyonaka 560-8531, Japan
}%

% \date{\today}
	
\begin{abstract}
 We propose a unifying framework for the state preparation using quantum power method algorithms based on generalized quantum signal processing (GQSP). We apply GQSP to realize quantum analogs of classical power iteration, power Lanczos, inverse iteration, and folded spectrum methods, all within a single coherent framework. GQSP allows efficient realization of methods that require complex polynomials, while avoiding the limitations of approaches based on linear combinations of time-evolution operators. Our constructions, including a Trotter-decomposition-free quantum inverse iteration, achieve near-optimal query scaling, together with reduced qubit requirements. The same formalism yields a quantum folded spectrum method for excited state preparation that avoids explicitly forming powers of the Hamiltonian or performing variational optimization. We provide a theoretical analysis of success probabilities and resource scaling, and we validate the methods numerically using molecular Hamiltonians. The results show that quantum power Lanczos lowers the computational cost and provides robust convergence compared to naive quantum power iteration. Our findings reveal that GQSP-based implementations of power methods combine scalability, flexibility, and robust convergence, paving the way for practical initial state preparations on fault-tolerant quantum devices.
\end{abstract}

\maketitle

\section{Introduction} \label{section_introduction}
Driven by advances in hardware architecture and algorithmic design, quantum computing has developed rapidly. Quantum computers are widely believed to outperform classical computers in condensed matter physics, quantum chemistry, and materials science~\cite{Bauer2020, Motta2021}. Solving the eigenvalue problem is a key computational challenge in these domains.

The most direct method for obtaining eigenvalues on a quantum computer is the quantum phase estimation (QPE) algorithm~\cite{Lloyd1996, Aspuru2005}. Despite its conceptual elegance, QPE typically necessitates substantial computational resources and fault-tolerant quantum computers (FTQCs). Consequently, proposals to reduce the overhead~\cite{Ding2023, Lin2022} face the fundamental challenge of preparing a high-fidelity initial approximation of the target state.

The variational quantum eigensolver (VQE)~\cite{Peruzzo2014}, which uses a hybrid quantum-classical approach, is a widely used algorithm for preparing an approximate quantum state. The VQE offers the advantage of shallow circuits that are compatible with existing hardware~\cite{Taube2006, Kandala2017, McCaskey2019}. However, its reliance on classical optimization makes it vulnerable to issues such as barren plateaus~\cite{Wang2021}, which limit its scalability for more expressive ansatzes.

Imaginary-time evolution (ITE) offers another route for ground state preparation by approximating a non-unitary propagator~\cite{Motta2020}. Although the classical ITE has long been a workhorse for quantum many-body simulations, quantum adaptations still face challenges. Many implementations inherit Trotter errors of time-evolution~\cite{McArdle2019, Li2024, Yi2025}, high resource costs to estimate the quantum Fisher information matrix~\cite{Kolotouros2025}, or post-selection overheads~\cite{chan2023simulatingnonunitarydynamicsusing}. Furthermore, ITE is naturally tailored to find ground states, thereby limiting its versatility for broader classes of problems.

By contrast, power methods represent a more general and flexible family of iterative algorithms with a long history of use in classical numerical linear algebra. In addition to ground-state preparation, power methods include algorithms for computing excited states, inverting matrices, solving linear systems, and performing principal component analysis. Therefore, they are widely used in numerous areas, such as computer-aided engineering, finance, and machine learning. In quantum chemistry, power-iteration-based schemes have also been developed for vibrational Hamiltonians, including tensor-structured power iteration~\cite{Leclerc2014}, Chebyshev methods~\cite{Leclerc2017}, and inverse iteration~\cite{Kallullathil2021}, demonstrating that such iterative approaches can be effective even in high-dimensional many-body settings.

Recent studies have explored various ways of implementing quantum power methods. Techniques include linear combinations of unitaries (LCU) to implement powers of cosine transformed Hamiltonians~\cite{Yimin2019}, finite-difference approximations of time evolution operators~\cite{Kazuhiro2021}, direct encoding using rotational operators~\cite{Britt2024}, HHL-based approach~\cite{HHL2009, NGHIEM2023} and quantum singular value transformation (QSVT) \cite{QSVT, ko2024quantumrandompowermethod, Nghiem2025}. Inverse iteration schemes have also been implemented~\cite{Kyriienko2020, Cainelli2024, Childs2017, Yoshikura2023}. Krylov subspace methods~\cite{Motta2024, Stair2020, yoshioka2024diagonalization} extend these ideas to more sophisticated approaches. However, many of these implementations face practical limitations, including Trotter errors, high qubit costs, and restrictions to only real-valued polynomials.

This study demonstrates that, for state preparation, various power methods can be implemented in a unified way using generalized quantum signal processing (GQSP) \cite{Motlagh2024,yu2022powerlimitationssinglequbitnative,QPP} (or more specifically generalized quantum singular value transformation (GQSVT) \cite{GQSVT}). The key advantage of GQSP over QSP~\cite{QSP} is that it can encode any general complex polynomial without an additional overhead. This paves the way for the implementation of the power Lanczos method, which may require complex coefficients, as well as for the future development of Green's function and complex energy shifts in the inverse iteration method. Furthermore, the GQSP gives advantage over methods that use a linear combination of time-evolution operators. For instance, our implementation of the quantum inverse iteration is free of Trotter error and requires fewer qubits in numerical tests than existing approaches. 

In addition, using the same technique, we introduced the quantum folded spectrum method (QFSM) to prepare the excited state wave functions, if approximate energy estimation is available. This method was previously used in the VQE framework, which required the explicit construction of a Hamiltonian squared. This results in numerous terms being measured~\cite{foldedVQE}. GQSP allows us to avoid the explicit construction of the powers of the Hamiltonian and variational optimization.

Throughout this work, we assume a fault-tolerant quantum computing setting, as the techniques we analyze require coherent implementation of polynomial transformations and precise phase operations. 
We provide a theoretical analysis of these quantum power methods, including estimates of success probability, query complexity, and qubit requirements. We show that the resulting algorithms achieve near-optimal scaling. 
We also construct explicit quantum circuits for each method and numerically validate their performance. 
In addition, we examine practical resource requirements, such as the total number of queries and qubits needed in realistic settings. 
Importantly, the proposed formulations remain mathematically equivalent to their classical counterparts; they can be simulated classically given exponential memory, and we confirm the expected convergence behavior.

The remainder of this paper is organized as follows. Section~\ref{sec:theory} reviews the fundamentals of GQSP and presents our implementation of various power methods. 
Section~\ref{sec:computation} provides computational details and a link to the reference code. 
Section~\ref{sec:results} presents theoretical analysis of resource scalings, followed by numerical demonstrations of the algorithms and comparisons with existing techniques, with particular attention to convergence behavior for molecular Hamiltonians and the associated computational costs.
Finally, Section~\ref{sec:conclusion} summarizes our findings, discusses potential improvements, and presents directions for future research and development.

\section{Theory} 
\label{sec:theory}

\subsection{Generalized quantum signal processing (GQSP)} 
\label{subsec:GQSP}
GQSP is a powerful framework that can be used for efficiently implementing polynomial transformations of a Hermitian operator~\cite{Motlagh2024}. It is first embedded in a larger unitary operator using a block encoding procedure, such as LCU~\cite{LCU}.

We seek to implement polynomial transformations of a Hamiltonian $H$. 
Within the block encoding framework, $H$ is rescaled as $\mathcal{H} = H/\lambda_\text{LCU}$, where $\lambda_\text{LCU}$ is the normalization constant associated with the LCU construction (see Appendix~\ref{appendix:block_encoding}).
The corresponding eigenvalues are therefore scaled as $\tilde{E}_i = E_i/\lambda_\text{LCU}$.

The LCU block encoding unitary acts, on each invariant subspace associated with an eigenvalue $\tilde{E}_i$, as
\begin{equation}
\label{eq:gqsp0}
U_\text{LCU} = 
\begin{pmatrix}
\mathcal{H} & \sqrt{I-\mathcal{H}^2}\\[1ex]
\sqrt{I-\mathcal{H}^2} & -\mathcal{H}
\end{pmatrix},
\end{equation}
which is a reflection.
For GQSP it is convenient to convert this reflection into a rotation by multiplying with a reflection about $|0\rangle$ on the ancilla:
\begin{equation}
\label{eq:gqsp00}
    U = (Z\otimes I)\, U_\text{LCU} = 
    \begin{pmatrix}
        \mathcal{H} & \sqrt{I-\mathcal{H}^2}\\[1ex]
        -\sqrt{I-\mathcal{H}^2} & \mathcal{H}
    \end{pmatrix},
\end{equation}
which acts as an $SU(2)$ rotation on each invariant subspace associated with $\tilde{E}_i$.

GQSP uses a $0$-controlled block encoding operator as the signal operator, denoted by $\operatorname{CU}_0$. This operator is defined as follows:
\begin{equation}
\label{eq:gqsp1}
\operatorname{CU}_0 = \bigl(|0\rangle\langle 0| \otimes U\bigr) + \bigl(|1\rangle\langle 1| \otimes I\bigr)
= \begin{pmatrix}
U & 0 \\
0 & I
\end{pmatrix}.
\end{equation}

To process the signal, GQSP employs general $SU(2)$ rotation operators on the auxiliary qubit. These rotations are defined as follows:
\begin{equation}
\label{eq:gqsp2}
R(\theta, \phi, \lambda)
= 
\begin{bmatrix}
e^{i(\lambda + \phi)} \cos(\theta) & e^{i\phi} \sin(\theta) \\[6pt]
e^{i\lambda} \sin(\theta)         & -\cos(\theta)
\end{bmatrix}
.
\end{equation}
Additionally, except for the first rotation, the parameter $\lambda$ can be set to zero without a loss of expressivity.

\begin{figure*}[ht]
  \centering
  \begin{adjustbox}{width=1\textwidth}
  \begin{quantikz}[column sep=5pt]
    \lstick{$\ket{0}$}
    & \gate{R(\theta_0, \phi_0, \lambda)}
    & \octrl{1}
    & \gate{Z}
    & \gate{R(\theta_1, \phi_1, 0)}
    & \octrl{1}
    & \gate{Z}
    & \ \ldots \
    & \gate{R(\theta_{d-1}, \phi_{d-1}, 0)}
    & \octrl{1}
    & \gate{Z}
    & \gate{R(\theta_{d}, \phi_{d}, 0)}
    & \\
    \lstick{$\ket{0}$}
    & \qwbundle{n}
    & \gate[2]{U_\text{LCU}}
    & \octrl{-1}
    &
    & \gate[2]{U_\text{LCU}}
    & \octrl{-1}
    & \ \ldots \
    &
    & \gate[2]{U_\text{LCU}}
    & \octrl{-1}
    &
    & \\
    \lstick{$\ket{\psi}$}
    & \qwbundle{m}
    &&&&&
    & \ \ldots \
    &&&&&
  \end{quantikz}
  \end{adjustbox}
    \caption{Generalized Quantum Signal Processing (GQSP) circuit. The circuit comprises three registers: the first one qubit register controls the GQSP procedure, the second $n$ qubits register performs the block encoding, and the third $m$ qubits register holds the state on which the Hamiltonian acts.}
  \label{fig:GQSP_cirsuit}
\end{figure*}

\begin{theorem}[Generalized Quantum Signal Processing \cite{Motlagh2024}]
\label{theorem:GQSP}
Let $U$ be the unitary operator. Then, there exist angles $\vec{\theta}=(\theta_0,\theta_1,\dots,\theta_d)$ and $\vec{\phi}=(\phi_0,\phi_1,\dots,\phi_d)$ in $\mathbb{R}^{d+1}$ along with a phase parameter $\lambda\in\mathbb{R}$ such that
\begin{equation}
\label{eq:gqsp3}
\begin{pmatrix}
P(U) & * \\
Q(U) & *
\end{pmatrix}
=
\left(\prod_{j=1}^d R(\theta_j,\phi_j,0)\; \operatorname{CU}_0\right)
R(\theta_0,\phi_0,\lambda),
\end{equation}
if and only if the following conditions hold:
\begin{enumerate}
    \item $P, Q \in \mathbb{C}[z]$ satisfying $\deg(P),\deg(Q) \le d$.
    \item $\forall z \in \mathbb{C}$ with $|z|=1$, the relation
    \[
    |P(z)|^2 + |Q(z)|^2 = 1
    \]
    holds.
\end{enumerate}
\end{theorem}
Here, the matrix elements denoted by * are determined implicitly by the unitarity of the full GQSP transformation.

It is essential for GQSP method to determine the angle sequences $\vec{\theta}$, $\vec{\phi}$, and the phase parameter $\lambda$ that produce the desired polynomial $P(z)$. As the first step, for any target polynomial $P(z)$, a complementary polynomial $Q(z)$ which satisfies $|P(z)|^2 + |Q(z)|^2 = 1$ for all $z$ on the complex unit circle must be constructed. It has been rigorously proven that this complementary polynomial exists for any $P(z)$ with $|P(z)| \le 1$ for all $z$ on the complex unit circle \cite{Motlagh2024}. Various methods exist for determining the appropriate angles, and further details are provided in the Appendix~\ref{appendix:angles}.

In a practical GQSP implementation, the LCU-to-qubitization reflection (see Eq.~\eqref{eq:gqsp00}) can be absorbed into the GQSP sequence of Eq.~\eqref{eq:gqsp3}. 
This is achieved by inserting a controlled-$Z$ gate whose target is the signal-processing ancilla and whose control is the $|0\rangle$ state of the LCU ancilla, effectively multiplying $U_{\text{LCU}}$ by the required reflection. 
The resulting quantum circuit is illustrated in Fig.~\ref{fig:GQSP_cirsuit}, and its structure is analogous to the GQSVT construction in Ref.~\cite{GQSVT}.

The next task involves obtaining a polynomial function of $\mathcal{H}$ from a polynomial function of $U$. This is accomplished by noting that the powers of $U$ generate Chebyshev polynomials of $\mathcal{H}$:
\begin{equation}
\label{eq:gqsp5}
U^k = 
\begin{pmatrix}
T_k(\mathcal{H}) & \ast\\[1ex]
\ast & \ast
\end{pmatrix},
\end{equation}
where $T_k(\mathcal{H})$ denotes the $k$-th Chebyshev polynomial evaluated at $\mathcal{H}$. 
This identity is a standard consequence of the qubitization construction: the unitary $U$ has eigenvalues $e^{\pm i\arccos(\tilde{E_i})}$ for each eigenvalue $\tilde{E_i}$ of $\mathcal{H}$, so $U^{k}$ implements $\cos\!\big(k\,\arccos(\tilde{E_i})\big) = T_{k}(\tilde{E_i})$.
Consequently, the operator $P(U)$ constructed via the GQSP block encodes a linear combination of Chebyshev polynomials of $\mathcal{H}$, which in turn can represent any polynomial function of $\mathcal{H}$.

\subsection{Quantum power iteration method (QPI)} 
\label{subsec:QPI}
Power iteration is a well-known technique for determining the largest magnitude eigenvalue of a matrix. We use this approach to find a ground state of the Hamiltonian $\mathcal{H}$. The underlying idea is straightforward: one chooses an initial guess for the eigenvector $|\psi^{(0)}\rangle$, which has a nonzero overlap with the true ground state, assuming that the ground state energy $\tilde{E}_0$ has the largest absolute value among all the eigenvalues of $\mathcal{H}$. The iterative procedure is defined as follows:

\begin{equation}
\label{eq:qpi1}
    |\psi^{(n+1)}\rangle = \frac{\mathcal{H} |\psi^{(n)}\rangle}{||\mathcal{H} |\psi^{(n)}\rangle||_2}.
\end{equation}

The error of this method scales as $O(|\tilde{E}_1/\tilde{E}_0|^n)$, where $\tilde{E}_2$ is the second largest eigenvalue in magnitude. Consequently, a larger spectral gap between $\tilde{E}_0$ and $\tilde{E}_1$ generally improves the convergence.

Equation~\eqref{eq:qpi1} reveals that the $n$-th iteration can be viewed as applying a polynomial of $\mathcal{H}$ to the initial state:

\begin{equation}
\label{eq:qpi2}
    |\psi^{(n)}\rangle = \frac{P(\mathcal{H}) |\psi^{(0)}\rangle}{||P(\mathcal{H}) |\psi^{(0)}\rangle||_2} = \frac{\mathcal{H}^n |\psi^{(0)}\rangle}{||\mathcal{H}^n |\psi^{(0)}\rangle||_2}.
\end{equation}

This polynomial transformation can be implemented using the GQSP algorithm. In particular, the powers of $\mathcal{H}$ can be expressed as linear combinations of Chebyshev polynomials. The general formula is as follows:

\begin{equation}
\label{eq:qpi5}
\begin{aligned}
\mathcal{H}^n
=\;& 2^{-n} \binom{n}{\tfrac{n}{2}}\, T_0(\mathcal{H}) \\[4pt]
&+ 2^{1-n}
   \sum_{\substack{k=1 \\ k \equiv n \,(\mathrm{mod}\, 2)}}^{n}
   \binom{n}{\tfrac{n-k}{2}}\, T_k(\mathcal{H}).
\end{aligned}
\end{equation}
where the first term contributes only when $n$ is even, and $T_k(\mathcal{H})$ denotes the $k$-th Chebyshev polynomial of $\mathcal{H}$.

The main disadvantage of this power method and all other methods presented in this study is that its success probability can be very small. It equals the square of the denominator in Eq.~\eqref{eq:qpi2}. A higher probability of success can be achieved with a better initial guess wave function $|\psi^{(0)}\rangle$. However, amplitude amplification is generally required, which deepens the circuit (see Sec.~\ref{subsec:theory_scaling}).

\subsection{Quantum power Lanczos method (QPL)} 
\label{subsec:QPL}
The power Lanczos (PL) method is a modification of the standard power iteration approach aimed at accelerating convergence by improving the initial guess for the eigenvector using Lanczos polynomials~\cite{Power_Lanczos}. The Lanczos algorithm constructs a Krylov subspace and determines an approximate ground-state eigenvector by diagonalizing a tridiagonal projection of the original Hamiltonian. More rapid convergence can be achieved compared with a naive power iteration by treating this approximate ground state as the initial vector in subsequent power iterations.

The classical Lanczos algorithm is a well-known technique used for computing several of the largest absolute eigenvalues in a matrix. To accomplish this, the Hamiltonian $\mathcal{H}$ is projected onto the Krylov subspace $\mathcal{K}_d(\mathcal{H},|\psi^{(0)}\rangle)$, which can be defined as follows:

\begin{equation}
\label{eq:qpl1}
\begin{aligned}
\mathcal{K}_d(\mathcal{H},|\psi^{(0)}\rangle)
= \operatorname{span}\bigl\{
&|\psi^{(0)}\rangle,\,
\mathcal{H}|\psi^{(0)}\rangle,\,
\mathcal{H}^2|\psi^{(0)}\rangle,\\
&
\ldots,\,
\mathcal{H}^{d-1}|\psi^{(0)}\rangle
\bigr\}.
\end{aligned}
\end{equation}
where $d$ is the dimension of the subspace and $|\psi^{(0)}\rangle$ is the initial state. Numerically stable implementations of the Lanczos algorithm use an orthonormal basis for this subspace. One standard method to build an orthonormal basis is through recurrence:

\begin{equation}
\label{eq:qpl2}
    \begin{aligned}
        &|\phi^{(0)}\rangle = \mathcal{H}|\psi^{(0)}\rangle - a_0 |\psi^{(0)}\rangle, \\
        &|\phi^{(n)}\rangle = \mathcal{H}|\psi^{(n)}\rangle - a_n |\psi^{(n)}\rangle - b_n |\phi^{(n-1)}\rangle,  
    \end{aligned}
\end{equation}
where $|\phi^{(n)}\rangle = b_{n+1} |\psi^{(n+1)}\rangle$ form a new orthonormal basis, $a_n = \langle \psi^{(n)} | \mathcal{H} | \psi^{(n)}\rangle$ and $b_n = \sqrt{ \langle \phi^{(n-1)}|\phi^{(n-1)}\rangle }$.

Thus, the matrix representation of Hamiltonian becomes tridiagonal, drastically simplifying its partial diagonalization. The eigenvalues and eigenvectors of this tridiagonal subproblem approximate those of the original matrix $\mathcal{H}$. The approximation can be systematically improved by increasing the dimensions of the Krylov subspace $d$; however, the rate and quality of convergence also depend on the distribution of the eigenvalues and their relative gaps.

While the standard Lanczos method provides a direct method to compute the tridiagonal representation and its eigendecomposition, it can be advantageous for treating the coefficients in the resulting Krylov vector expansion as variable parameters. In this approach, the energy $\langle \Psi_k^{(0)} \lvert \mathcal{H} \rvert \Psi_k^{(0)} \rangle$ is minimized in trial state $| \Psi_k^{(0)} \rangle$ in the Krylov subspace. The key advantage of this approach lies in it eliminating the need to precisely calculate the matrix elements of the effective Hamiltonian in the Krylov space. For the ground state problem, this minimization reproduces the usual Lanczos result; however, it can offer improved numerical stability.

For instance, consider a Krylov subspace with dimension $k+1$. A generic trial state then takes the following form:
\begin{equation}
\label{eq:qpl3}
    \begin{aligned}
        &|\Psi_1^{(0)}\rangle = |\psi^{(0)}\rangle + C_1 \mathcal{H} |\psi^{(0)}\rangle, \\
        &|\Psi_2^{(0)}\rangle = |\psi^{(0)}\rangle + C_1 \mathcal{H} |\psi^{(0)}\rangle + C_2 \mathcal{H}^2 |\psi^{(0)}\rangle,  
    \end{aligned}
\end{equation}
and so on. The coefficients $C_i$ can be determined by solving the linear equations derived from the tridiagonal matrix or, as advocated here, by performing a variational minimization of the energy. This optimization involves only a small number of parameters since $k$ is often small; therefore, it avoids most of the problems related to VQE.

Once the optimized Krylov (Lanczos) state $| \Psi_k^{(0)} \rangle$ is determined, it is used as a more refined starting point for power iteration; that is, instead of $|\psi^{(0)}\rangle$, we initialize our iterative procedure with $|\Psi_k^{(0)}\rangle$. The subsequent $n$-th power iteration state can then be written as follows:
\begin{equation}
\label{eq:qpl4}
\begin{aligned}
    |\Psi_k^{(n)}\rangle = &\frac{1}{N} \mathcal{H}^n |\Psi_k^{(0)}\rangle  \\
    &=\frac{1}{N} \mathcal{H}^n (1+C_1\mathcal{H} + \cdots + C_k \mathcal{H}^k) |\psi^{(0)}\rangle,
\end{aligned}
\end{equation}
where $N$ is a normalization factor. We will call this method QPL$k$, specifying the degree of the Lanczos polynomial. The polynomial acting on the states can be naturally implemented within the GQSP framework. Here, GQSP demonstrates its advantage over conventional QSP, since the parametric $C_i$ can become complex during optimization in order to improve convergence. Although the coefficients are ultimately real when calculated without considering the magnetic interaction, entering the complex plane in this manner helps finding a workaround to the minimum.

In practice, the QPL method proceeds in two main steps:
\begin{enumerate}
    \item 
    \textbf{Variational Lanczos state preparation.} 
    Build $| \Psi_k^{(0)}\rangle$ using a polynomial in $\mathcal{H}$ (truncated at order $k$) using GQSP, then optimize the coefficients $\{C_i\}$ to minimize $\langle \Psi_k^{(0)} | \mathcal{H} | \Psi_k^{(0)}\rangle$.
    While one can solve for $\{C_i\}$ analytically using the Lanczos tridiagonalization, this requires computing expectation values of multiple powers of $\mathcal{H}$. In many quantum settings, a small variational optimization with a handful of coefficients is simpler.
    
    \item 
    \textbf{Quantum power iteration.} 
    After obtaining $| \Psi_k^{(0)}\rangle$, perform the standard power iteration 
    (see Eq.~\eqref{eq:qpl4}) 
    using this improved initial guess. 
\end{enumerate}

We further show that even a small Krylov subspace ($k=1$ or $k=2$) is often sufficient for providing a significantly better initial state. A larger $k$ yields a more difficult optimization landscape but often offers insignificant gains in terms of accuracy. Thus, the QPL method can serve as a middle ground: it leverages a minimal Lanczos subspace to accelerate convergence in the power iteration by introducing only a few parameters. This approach can noticeably outperform pure power iteration for ground state computations, particularly when the spectrum of $\mathcal{H}$ has a large gap or when a small Krylov space already captures a significant fraction of the ground state component in $| \psi^{(0)}\rangle$. 

Most other Lanczos-inspired algorithms for quantum computers rely on the explicit construction of the Lanczos tridiagonal matrix. 
For example, Kirby et al.~\cite{Kirby2023exactefficient} employ a Chebyshev-polynomial Krylov subspace, using a quantum device to evaluate matrix elements that are subsequently diagonalized on a classical computer. 
Baker~\cite{Baker2021} follows the standard Lanczos recursion to construct the Krylov basis and likewise performs explicit tridiagonalization.

In contrast, the QPL approach replaces explicit matrix construction and diagonalization with variational optimization of the Krylov-space coefficients. 
This is the central conceptual difference from other methods. 
By avoiding the calculation of effective Hamiltonian matrix elements in the Krylov basis, the method reduces numerical overhead and potential sources of instability. 
For ground state estimation, the variational minimization recovers the usual Lanczos result while providing improved numerical stability~\cite{Power_Lanczos}.

\subsection{Quantum inverse iteration method (QII)} 
\label{subsec:QII}
Inverse iteration is another classical technique used to solve eigenvalue problems iteratively. Unlike the power iteration, which is naturally suited for finding the dominant eigenvalue and its corresponding eigenvector, the inverse iteration can be used to approximate any eigenpair, provided that an appropriate shift is chosen. Even though in our implementation it can only find the ground state, having a good enough energy approximation, the inverse iteration can converge much faster than the power iteration. Specifically, one starts with an initial guess of the eigenvector $\lvert \psi^{(0)} \rangle$ and an approximate eigenvalue $\epsilon_0$ close to the true eigenvalue $\tilde{E}_0$. The iterative step is given by

\begin{equation}
\label{eq:qii1}
    |\psi^{(n+1)}\rangle = \frac{(\mathcal{H}-\epsilon_0 I)^{-1} |\psi^{(n)}\rangle}{||(\mathcal{H}-\epsilon_0 I)^{-1} |\psi^{(n)}\rangle||_2}.
\end{equation}

The convergence rate depends on how well $\epsilon_0$ approximates the target eigenvalue $\tilde{E}_0$. Quantitatively, if $\tilde{E}_1$ is the eigenvalue closest to $\epsilon_0$ other than $\tilde{E}_0$, the error drops as $O\bigg(\biggl|(\epsilon_0 - \tilde{E}_0)/(\epsilon_0 - \tilde{E}_1)\biggr|^n\bigg)$.
Hence, choosing $\epsilon_0$ closer to $\tilde{E}_0$ accelerates the convergence.

By the repeated application of $(\mathcal{H} - \epsilon_0 I)^{-1}$, we see that the $n$-th iterate can be written as follows:
\begin{equation}
\label{eq:qii2}
    |\psi^{(n)}\rangle = \frac{(\mathcal{H}-\epsilon_0 I)^{-n} |\psi^{(0)}\rangle}{||(\mathcal{H}-\epsilon_0 I)^{-n} |\psi^{(0)}\rangle||_2}.
\end{equation}

A polynomial approximation of the operator $(\mathcal{H} - \epsilon_0 I)^{-n}$ is required to implement this step using the GQSP framework. The Taylor power series expansion for the negative powers of a shifted operator is given by
\begin{equation}
\label{eq:qii3}
    (\mathcal{H}-\epsilon_0 I)^{-n} = (-1)^n \sum_{k=0}^{\infty} \binom{n+k-1}{k} \ \frac{1}{\epsilon_0^{n+k}} \mathcal{H}^k.
\end{equation}

In principle, an infinite series can be truncated to obtain a polynomial in $\mathcal{H}$. The radius of convergence for this expansion is $\| \mathcal{H} \|_2 < |\epsilon_0|$. In practice, this condition can be restrictive for excited state calculations because we require $\epsilon_0$ to be as close as possible to $\tilde{E}_0$. We need $\epsilon_0$ to be outside the spectral radius of $\mathcal{H}$, that is, $\| \mathcal{H} \|_2 < |\epsilon_0|$, to guarantee the convergence of the series. This is naturally satisfied if $\tilde{E}_0$ is the ground-state energy and $\epsilon_0$ is chosen to be slightly smaller than $\tilde{E}_0$. In this scenario, one can often achieve a more rapid convergence compared to QPI, owing to the favorable shift.

Similarly to the power iteration method, a key step in QII involves mapping the truncated power series onto a polynomial expansion in terms of Chebyshev polynomials, which are naturally compatible with QSP. Once the polynomial representation is finalized, the same QSP techniques used for power iteration can be used to apply $(\mathcal{H} - \epsilon_0 I)^{-n}$ to a quantum state.

The full power of GQSP became apparent when a complex energy shift $z = \epsilon_0 + i\zeta$ was introduced. While we did not employ it in the examples below, this shift regularizes the pole at $E = \epsilon_0$ and enlarges the radius of convergence of the Taylor power series expansion by displacing the singularities off the real axis. Specifically, the shifted resolvent $(\mathcal{H} - z I)^{-1}$ remains finite at $E = \epsilon_0$ and, in the spectral representation, its squared magnitude takes the Lorentzian form $\left|(\mathcal{H} - z I)^{-1}\right|^2 = \left((\mathcal{H} - \epsilon_0 I)^2 + \zeta^2 I\right)^{-1}$. The width parameter $\zeta$ governs the trade-off between spectral selectivity (a narrow peak for a small $\zeta$) and numerical stability or convergence speed (a larger $\zeta$ produces better conditioning). Finally, note that $(\mathcal{H} - z I)^{-1}$ is precisely the Green’s function of $\mathcal{H}$ evaluated at $z$, establishing a direct connection to the contour integral and filter diagonalization eigensolvers and other Green’s function methods.

\subsection{Quantum folded spectrum method (QFSM)} 
\label{subsec:QFSM}
An iterative version of the folded spectrum method can be employed to compute the excited states within the GQSP framework. The key concept was the construction of a modified operator:
\begin{equation}
\label{eq:qfsm1}
    \mathcal{H}' \;=\; I \;-\; C\,\bigl(\mathcal{H} - \epsilon_m I\bigr)^2,
\end{equation}
where $\epsilon_m$ is an estimate of the target eigenvalue and $C$ is an appropriately chosen normalization constant. By focusing on $\bigl(\mathcal{H}-\epsilon_m I\bigr)^2$, one effectively "folds" the spectrum around $\epsilon_m$. After scaling with $C$ and shifting by identity, the eigenvalue closest to $\epsilon_m$ in the original spectrum becomes the largest eigenvalue of $\mathcal{H}'$; thus, this can be achieved via a power iteration scheme:

\begin{equation}
\label{eq:qfsm2}
    |\psi_m^{(n+1)}\rangle = \frac{(I - C(\mathcal{H}-\epsilon_m I)^2) |\psi_m^{(n)}\rangle}{||(I - C(\mathcal{H}-\epsilon_m I)^2) |\psi_m^{(n)}\rangle||_2}.
\end{equation}

Repeating this step $n$ times yields
\begin{equation}
\label{eq:qfsm3}
    |\psi_m^{(n)}\rangle = \frac{(I - C(\mathcal{H}-\epsilon_m I)^2)^n |\psi_m^{(0)}\rangle}{||(I - C(\mathcal{H}-\epsilon_m I)^2)^n |\psi_m^{(0)}\rangle||_2}.
\end{equation}
Because $\bigl(I - C(\mathcal{H}-\epsilon_m I)^2\bigr)^n$ is a polynomial in $\mathcal{H}$, it is naturally suited for GQSP implementations. In practice, the Hamiltonian is first shifted by replacing $\mathcal{H}$ with $\mathcal{H}_{\text{shifted}} = \mathcal{H} - \epsilon_m I$. Specifically, shifting alters only the identity term in the LCU construction. After this shift, the polynomial 
\begin{equation}
\label{eq:qfsm4}
    (1 - C \mathcal{H}_{\text{shifted}}^2)^n = \sum^{n}_{k=0} \binom{n}{k} (-C)^k \mathcal{H}^{2k}_{\text{shifted}}
\end{equation}
can be applied exactly to the quantum state. However, the convergence in this folded spectrum scheme depends sensitively on the choice of $\epsilon_m$, the initial guess $| \psi_m^{(0)}\rangle$, and the constant $C$. In typical implementations, $C$ is chosen such that the shifted square spectrum lies between $0$ and $1$, which facilitates a faster convergence. However, the initial state can have an even more profound effect on the performance, as illustrated in the next section.

Table~\ref{tab:methods} summarizes the polynomial forms of all the methods discussed in this study.

\begin{table}[h]
    \centering
    \begin{tabular}{l l}
        \hline\hline
        \textbf{Method} & \textbf{Polynomial Form} \\
        \hline\hline
        QPI 
        & 
        \(\displaystyle \mathcal{H}^n\)
        \\
        \\
        QPL$k$ 
        & 
        \(\displaystyle 
          \mathcal{H}^n \,\sum_{i=0}^{k} C_i\,\mathcal{H}^i
          \)
        \\
\\
        QII 
        & 
        \(\displaystyle 
          (-1)^n \sum_{k=0}^{\infty} 
          \binom{n+k-1}{k} \,\frac{1}{\epsilon_0^{n+k}}\, \mathcal{H}^k
          \)
        \\
\\
        QFSM 
        & 
        \(\displaystyle 
          \sum_{k=0}^{n} 
          \binom{n}{k} \,(-C)^k\,
          \mathcal{H}_{\text{shifted}}^{\,2k}
         \), \ 
        where 
        $\mathcal{H}_{\text{shifted}} = \mathcal{H} - \epsilon_m I$
        \\
        \hline\hline
    \end{tabular}
    \caption{Summary of the polynomial expansions that underlie each quantum iterative method (QPI, QPL, QII, and QFSM). Here, $\epsilon_m$ is the target eigenvalue approximation, and $C$ is chosen to ensure the folded spectrum operator is suitably bounded.}
    \label{tab:methods}
\end{table}

\section{Computational details} 
\label{sec:computation}

Here, we provide a computational setup for proof-of-concept numerical simulations of our proposed methods. The reference CASCI calculations were performed using the PySCF package~\cite{Sun2020}. The Jordan–Wigner transformed molecular Hamiltonians used in our simulations were generated using the Quket package~\cite{quket}.
The QSub subpackage of the QURI Parts framework~\cite{quri_parts} was used for resource estimation and fault-tolerant circuit compilation.

Except for the QFSM, all simulations employed the qubit tapering technique to reduce the number of qubits~\cite{bravyi2017taperingqubitssimulatefermionic, Setia2020}. For instance, the H$_2$ molecule in a $(2, 2)$ active space only required $1$ qubit instead of $4$, CH$_2$ in a $(6, 6)$ active space required $8$ qubits instead of $12$, and N$_2$ in a $(6, 6)$ space required $7$ qubits instead of $12$.

The GQSP circuit illustrated in Fig.~\ref{fig:GQSP_cirsuit} was implemented using the Qulacs simulator~\cite{Suzuki2021qulacsfast}. The angle estimation was performed using Prony’s method~\cite{yamamoto2024robustanglefindinggeneralized}, as described in detail in Appendix~\ref{appendix:angles}. The \textit{capitalization} step required an additional qubit.

Block encoding was implemented using the LCU method only for the H$_2$ system, which required two additional qubits. To maintain the computational feasibility for the other systems, an explicit block encoding scheme was employed using a single additional qubit, as illustrated in Eq.~\eqref{eq:gqsp00}. While in practical implementation on a quantum device the LCU is necessary, we used this technique for performance verification purposes. Further details are provided in Appendix~\ref{appendix:angles}. The complete implementation is available on our GitHub repository: \url{https://github.com/mizukami-group/power-lanczos-data-FY2025.git}.

\section{Results and Discussion} 
\label{sec:results}

\subsection{Theoretical scaling}
\label{subsec:theory_scaling}

Let $\mathcal{H}$ be a Hamiltonian with eigenpairs $\{(\tilde{E}_i,\lvert \tilde{E}_i\rangle)\}_{i=0}^{n}$ ordered as $\tilde{E}_0<\tilde{E}_1<\cdots<\tilde{E}_n$. 
Let the spectral gap be $\tilde{\Delta}:=\tilde{E}_1-\tilde{E}_0>0$. 
We assume that $\tilde{E}_0$ and $\tilde{E}_1$ are negative and have the largest absolute values among the spectrum, and define their ratio as $\rho := \tilde{E}_1/\tilde{E}_0$, which satisfies $0 < \rho < 1$.
Assume that $\tilde{\Delta} \ll |\tilde{E}_0|$ so that $\log(1/\rho)\approx \tilde{\Delta}/|\tilde{E}_0|$.

GQSP implements a polynomial filter $P(\mathcal{H})$ on an initial state $\lvert\psi^{(0)}\rangle$,
\begin{equation}
    \lvert\psi^{(0)}\rangle \longmapsto \frac{P(\mathcal{H})\lvert\psi^{(0)}\rangle}{\lVert P(\mathcal{H})\lvert\psi^{(0)}\rangle\rVert_2},
\end{equation}
with postselection success probability
\begin{equation}
\begin{aligned}
p_{\mathrm{succ}}
&= \bigl\lVert P(\mathcal{H})\lvert\psi^{(0)}\rangle \bigr\rVert_2^2 \\
&= \langle\psi^{(0)}\rvert\, P(\mathcal{H})^{\dagger}P(\mathcal{H})\,\lvert\psi^{(0)}\rangle.
\end{aligned}
\end{equation}

Amplitude amplification (AA) increases this probability to 
\begin{equation}
    p^{(m)}_{\mathrm{succ}}=\sin^2((2m+1)\theta)
\end{equation}
with $\theta=\arcsin\sqrt{p_{\mathrm{succ}}}$. 
Choosing
\begin{equation}
    m=\Bigl\lfloor \frac{\pi}{4\theta}-\frac12\Bigr\rfloor \approx \frac{\pi}{4\sqrt{p_{\mathrm{succ}}}}-\frac12
\end{equation}
maximizes the success probability. 
Each Grover step uses the underlying block a constant number of times, so the cost scales by $(2m+1)$.

\textbf{QPI.}
The initial guess wave function can be written as $\lvert\psi^{(0)}\rangle=\sum_{i=0}^{n} a_i \lvert \tilde{E}_i\rangle$, where we assume that the ground state contribution is lower bounded $|a_0| \ge \gamma$. 
After the $k$ QPI steps, the ground state overlap is
\begin{equation}
\label{eq:QPI_overlap}
\begin{aligned}
p_k \;=\;&\; \frac{|a_0|^2}{|a_0|^2+\sum_{i>0}|a_i|^2\,(\tilde{E}_i/\tilde{E}_0)^{2k}} \\[4pt]
\ge&\; \frac{1}{1+\frac{1-|a_0|^2}{|a_0|^2}\,\rho^{2k}}.
\end{aligned}
\end{equation}

The postselection success probability satisfies
\begin{equation}
\label{eq:QPI_succ}
    p^{(k)}_{\mathrm{succ}}
\;\ge\;
\frac{\bigl(|a_0|^2-(1-|a_0|^2)\rho^{k}\bigr)^{2}}{|a_0|^2+(1-|a_0|^2)\rho^{2k}},
\end{equation}
so as $k\to\infty$ we have $p_k\to 1$ and $p^{(k)}_{\mathrm{succ}}\to |a_0|^{2}$. 
With AA and $|a_0|\ge\gamma$, the number of amplification steps can be chosen as $m_\star\approx \pi/(4\gamma)-1/2$.
Since degree $k$ requires $k$ queries to block encoding, this is multiplied by $(2m_\star+1)$, which gives about $\pi k/(2\gamma)$ of queries to block encoding to implement QPI with AA.

To reach an overlap with the ground state $(1-\varepsilon)$, a sufficient condition can be derived from Eq.~\eqref{eq:QPI_overlap}:
\begin{equation}
\begin{aligned}
k \;\ge\;&\; \frac{1}{2\log(1/\rho)}\Bigl[\log\frac{1}{\varepsilon}+2\log\frac{1}{\gamma}\Bigr] \\[4pt]
=\;&\; \frac{1}{2\log(1/\rho)}\log\frac{1}{\gamma^2\varepsilon}.
\end{aligned}
\end{equation}

Using $\log(1/\rho)\approx \tilde{\Delta}/|\tilde{E}_0|$, the resulting query complexity is
\begin{equation}
    \text{Queries}_{\mathrm{QPI+AA}}
\;=\;
\frac{\pi}{4}\,\frac{1}{\tilde{\Delta}\,\gamma}\,\log\frac{1}{\gamma^{2}\varepsilon}.
\end{equation}
This is a near-optimal scaling for state preparation without energy estimation~\cite{lin2020}.

The QPI prepared state can be used for QPE to obtain the ground state energy.
We assume that the prepared state is almost exact, so no further AA is required for QPE.
The qubitization implementation of QPE to estimate the energy up to precision $\tilde{\varepsilon}_E$ requires the following number of block encoding queries~\cite{Lee2021}:
\begin{equation}
    \text{Queries}_{\mathrm{QPE}}=\left\lceil \frac{\pi}{2\,\tilde{\varepsilon}_E}\right\rceil.
\end{equation}

Thus, when the energy variables are returned from the rescaled Hamiltonian $\mathcal{H} = H/\lambda_\text{LCU}$ to the original units, the number of queries becomes
\begin{equation}
\begin{aligned}
\text{Queries}_{\mathrm{QPI+AA+QPE}}
\;=\;&\;
\frac{\pi}{4}\,\frac{\lambda_\text{LCU}}{\Delta\,\gamma}\,
\log\frac{1}{\gamma^{2}\varepsilon} \\[4pt]
&\;+\;
\left\lceil \frac{\pi\,\lambda_\text{LCU}}{2\,\varepsilon_E}\right\rceil.
\end{aligned}
\end{equation}
and if one sets $\varepsilon_E=\Theta(\Delta)$ to resolve the ground energy within the gap, the second term is $\Theta(\lambda_\text{LCU}/\Delta)$.

\textbf{QPL.}
QPL optimizes the polynomial filter over the Krylov subspace. 
This modification does not influence the scaling, but changes the prefactor. 
We model the effect of a Lanczos step as increasing the ground state amplitude $a_0$ by a factor $\beta \ge 1$, so that the effective lower bound becomes $\beta\gamma$.
As a result,
\begin{equation}
\label{eq:QPL_queries}
\text{Queries}_{\mathrm{QPL+AA}}
\;=\;
\frac{\pi}{4}\,\frac{\lambda}{\Delta\,\gamma}\cdot \frac{1}{\beta}\;
\log\!\frac{1}{\beta^2\gamma^{2}\varepsilon}.
\end{equation}
In practice, this constant-factor improvement reduces iteration counts; in our numerical simulations shown in the next section, it saves up to roughly $50$ iterations.

\begin{table*}[t]
\centering
\renewcommand{\arraystretch}{1.2}
\begin{tabular}{ll}
\hline\hline
\textbf{Method / task} & \textbf{Queries to block encoding} \\
\hline\hline
\multicolumn{2}{l}{\emph{This work (GQSP polynomials with AA)}}\\
QPI + AA (ground state, overlap $1-\varepsilon$) &
$\displaystyle \frac{\pi}{4}\,\frac{\lambda}{\Delta\,\gamma}\,
\log\!\frac{1}{\gamma^{2}\varepsilon}$ \\[4pt]
QPI + AA + QPE (energy to precision $\varepsilon_E$) &
$\displaystyle \frac{\pi}{4}\,\frac{\lambda}{\Delta\,\gamma}\,
\log\!\frac{1}{\gamma^{2}\varepsilon}
\;+\;
\Big\lceil \frac{\pi\,\lambda}{2\,\varepsilon_E}\Big\rceil$ \\[4pt]
QPL + AA (ground state, overlap $1-\varepsilon$) &
$\displaystyle \frac{1}{\beta}\cdot
\frac{\pi}{4}\,\frac{\lambda}{\Delta\,\gamma}\,
\log\!\frac{1}{\beta^2\gamma^{2}\varepsilon}$ \quad ($\beta>1$) \\[4pt]
QII + AA (ground state, overlap $1-\varepsilon$) &
$\displaystyle O\!\Biggl(
\frac{\lambda_\text{LCU}}{\gamma\,(\Delta-\varepsilon_E)}
\max  \Biggl\{ \frac{\log\!\bigl(1/(\gamma^{2}\varepsilon)\bigr)}
{\log\!\bigl((\Delta-\varepsilon_E)/\varepsilon_E\bigr)}, \log \frac{1}{\varepsilon} \Biggr\}
\Biggr)$\\
& $\displaystyle \Theta\Bigg(\frac{\lambda}{\gamma \Delta}\log\frac{1}{\varepsilon}\Bigg)$ \quad ($\varepsilon_E = 0$)\\[4pt]
QFSM + AA (excited state, overlap $1-\varepsilon$) &
$\displaystyle O\!\Bigg(
\frac{\lambda^2}{\gamma\,\Delta(\Delta-2\varepsilon_E)}\,
\log\!\frac{1}{\gamma^{2}\varepsilon}
\Bigg)$\\[4pt]
\hline\hline
\multicolumn{2}{l}{\emph{Literature baselines}}\\
Lin--Tong (2020): ground state, $E_0$ bound known &
$\displaystyle O\!\Big(
\frac{\lambda}{\gamma\,\Delta}\,
\log\!\frac{1}{\varepsilon}
\Big)$ \,{\small\cite{lin2020}} \\[4pt]
Lin--Tong (2020): ground energy to precision $\varepsilon_E$ &
$\displaystyle \tilde O\!\Big(
\frac{\lambda}{\gamma\,\varepsilon_E}\,
\log\!\frac{1}{\vartheta}
\Big)$ \,{\small\cite{lin2020}} \\[4pt]
Lin--Tong (2020): ground state, $E_0$ bound unknown &
$\displaystyle \tilde O\!\Big(
\frac{\lambda}{\gamma\,\Delta}\,
\log\!\frac{1}{\vartheta\,\varepsilon}
\Big)$ \,{\small\cite{lin2020}} \\[4pt]
Ge--Tura--Cirac (2019): ground state, $E_0$ well known &
$\displaystyle \tilde O\!\Big(
\frac{\lambda}{\gamma\,\Delta}
\Big)$ \,{\small\cite{ge2019}} \\[4pt]
Ge--Tura--Cirac (2019): ground energy to precision $\varepsilon_E$ &
$\displaystyle \tilde O\!\Big(
\frac{\lambda^{3/2}}{\gamma\,\varepsilon_E^{3/2}}
\Big)$ \,{\small\cite{ge2019}} \\[4pt]
Ge--Tura--Cirac (2019): ground state, $E_0$ bound unknown &
$\displaystyle \tilde O\!\Big(
\frac{\lambda^{3/2}}{\gamma\,\Delta^{3/2}}
\Big)$ \,{\small\cite{ge2019}} \\
\hline\hline
\end{tabular}
\caption{Query complexities, stated in a unified notation.
Here $\lambda$ denotes the block encoding normalization (corresponding to $\alpha$ in Refs.~\cite{lin2020,ge2019}), $\Delta$ the energy gap, $\gamma$ a lower bound on the initial overlap, $\varepsilon$ the state-preparation error, $\varepsilon_E$ the energy estimation precision (denoted $h$ in Ref.~\cite{lin2020}), and $\vartheta$ a failure-probability parameter.
Our methods use GQSP polynomials combined with AA. 
In the literature block we retain the original authors' asymptotic notation $O(\cdot)$, $\tilde O(\cdot)$ as written in Refs.~\cite{lin2020,ge2019}. $\tilde{O}(\cdot)$ hides polylogarithmic factors in $1/\Delta$, $1/\varepsilon$, $1/\varepsilon_E$ and $1/\gamma$.}
\label{tab:query-comparison}
\end{table*}

\textbf{QII.}
When an estimate of the ground state energy is available, QII provides an efficient route for ground state preparation.  
Let $\mathcal{H}$ be a Hamiltonian with eigenpairs $(\tilde{E}_i,|\tilde{E}_i\rangle)$ ordered as $\tilde{E}_0 < \tilde{E}_1 < \dots < \tilde{E}_n$, and let $\epsilon_0$ be an approximation to the ground state energy satisfying $\lvert \epsilon_0 - \tilde{E}_0\rvert \le \tilde{\varepsilon}_E$.

Writing $|\psi^{(0)}\rangle = \sum_i a_i |\tilde{E}_i\rangle$ with $|a_0|\ge\gamma$, the overlap with the ground state after $k$ iterations is
\begin{equation}
\label{eq:QII_overlap}
p_k = \frac{|a_0|^2}{|a_0|^2 + \sum_{i>0}|a_i|^2 
\left|\frac{\tilde{E}_0-\epsilon_0}{\tilde{E}_i-\epsilon_0}\right|^{2k}}.
\end{equation}
Let
\begin{equation}
    \eta_{\mathrm{inv}} := \max_{i>0}\left|\frac{\tilde{E}_0-\epsilon_0}{\tilde{E}_i-\epsilon_0}\right| \le \frac{\tilde{\varepsilon}_E}{\Delta - \tilde{\varepsilon}_E}, \qquad \tilde{\Delta} = \tilde{E}_1 - \tilde{E}_0,
\end{equation}
assuming $\tilde{\varepsilon}_E < \tilde{\Delta}/2$.

Then the lower bound of the overlap becomes
\begin{equation}
\label{eq:QII_overlap_bound}
p_k \ge \frac{1}{1+\frac{1-|a_0|^2}{|a_0|^2}\,\eta_{\mathrm{inv}}^{2k}}.
\end{equation}
The success probability for QII is also analogous to the QPI formula with replacement of $\rho$ with $\eta_{\mathrm{inv}}$ (see Eq.~\eqref{eq:QPI_succ}).

To achieve overlap $(1-\varepsilon)$, the sufficient number of steps is 
\begin{equation}
\label{eq:QII_k_bound}
k \ge \frac{1}{2\log(1/\eta_{\mathrm{inv}})} \log\frac{1}{\gamma^2\varepsilon}.
\end{equation}
Thus, the number of required iterations decreases rapidly as the energy estimate $\tilde{\varepsilon}_E$ approaches zero.

In the GQSP framework, the inverse expression $f_k(\mathcal{H}) = (\mathcal{H} - \epsilon_0 I)^{-k}$ is implemented by a polynomial approximation $P_d(\mathcal{H})$ of degree $d$.
Since $\mathcal{H}$ is already the rescaled block-encoded Hamiltonian with spectrum contained in $[-1, 1]$, the only singularities of $f_k(x)$ arise at points where $x=\epsilon_0$.

Let $\tilde{g}:=\tilde{\Delta}-\tilde{\varepsilon}_E$.
Under the assumption $\tilde{\varepsilon}_E < \tilde{\Delta}/2$, the excited state eigenvalues $\tilde{E}_i$ satisfy $\lvert \tilde{E}_i-\epsilon_0\rvert\ge \tilde{g}$, so $f_k(x)$ is analytic on the subset of the spectrum obeying $|x-\epsilon_0| \ge \tilde{g}$.
In other words, the nearest singularity of $f_k(x)$ lies at distance $\tilde{g}$ from the spectral interval on which the approximation must hold.

For analytic functions on $[-1, 1]$ with nearest singularity at distance $\tilde{g}$, standard Chebyshev/Bernstein–ellipse bounds imply the existence of a polynomial $P_d$ satisfying
\begin{equation}
    \sup_{x:\, |x-\epsilon_0| \ge \tilde{g}} \bigl| f_k(x) - P_d(x) \bigr| \le \delta .
\end{equation}
whenever
\begin{equation}
\label{eq:QIIpoly_degree}
d = \Theta\left(\frac{1}{\tilde{g}}\Bigl[k + \log\frac{1}{\delta}\Bigr]\right).
\end{equation}
This equation gives the polynomial degree required to achieve operator norm error $\delta$ in a single-shot implementation of $f_k(\mathcal{H})$.

A single round of AA boosts postselection, costing a factor of $(2m_\star+1)$, where $m_\star\approx \pi/(4\gamma)-1/2$.
Choosing $\delta=\Theta(\varepsilon)$, so one application of polynomial, and returning back to the original units, gives the total query complexity
\begin{equation}
\label{eq:QIIpoly_queries_exact}
\text{Queries}_{\mathrm{QII+AA}}
\;=\;
\Theta\!\left(
\frac{1}{\gamma}\cdot
\frac{\lambda_\text{LCU}}{\Delta-\varepsilon_E}\,
\Bigl[k+\log\!\frac{1}{\varepsilon}\Bigr]
\right).
\end{equation}
Substituting \eqref{eq:QII_k_bound} for $k$,
\begin{equation}
\label{eq:QIIpoly_queries_final}
\begin{aligned}
\text{Queries}&_{\mathrm{QII+AA}}
\;=\;
O\!\Biggl(
\frac{\lambda_\text{LCU}}{\gamma\,(\Delta-\varepsilon_E)} \\
&\times
\max  \Biggl\{ \frac{\log\!\bigl(1/(\gamma^{2}\varepsilon)\bigr)}
{\log\!\bigl((\Delta-\varepsilon_E)/\varepsilon_E\bigr)}, \log \frac{1}{\varepsilon} \Biggr\}
\Biggr).
\end{aligned}
\end{equation}
If $\epsilon_0=\tilde{E}_0$ (i.e., $\varepsilon_E=0$), one can take $k=1$ in exact arithmetic; practically, the polynomial degree $d=\Theta((\lambda_\text{LCU}/\Delta)\log(1/\varepsilon))$ then dominates.

\textbf{QFSM.}
To prepare the excited state, QFSM can be used.
Given a target excited level $E_j$ and an energy estimate $\epsilon_j$ with $\lvert \epsilon_j-\tilde{E}_j\rvert\le \tilde{\varepsilon}_E$, QFSM applies the polynomial filter $F_k(\mathcal{H})\;=\;\bigl(I - C\,(\mathcal{H}-\epsilon_j I)^2\bigr)^{k}$, which concentrates weight near the eigenvalue closest to $\epsilon_j$.
Choose $C$ so that $0\le 1- C (\tilde{E}_i-\epsilon_j)^2\le 1$ on the spectrum; a sufficient choice is
\begin{equation}
    0<C\le \frac{1}{\tilde{R}^2},\qquad \tilde{R}\;\ge\; \max_{i}\lvert \tilde{E}_i-\epsilon_j\rvert.
\end{equation}

Write $\lvert\psi^{(0)}\rangle=\sum_{i=0}^{n} a_i \lvert \tilde{E}_i\rangle$ and assume $\lvert a_j\rvert\ge \gamma$ for the target state $\lvert \tilde{E}_j\rangle$.
Let 
\begin{equation}
\begin{aligned}
    \tilde{g}\;:=\; \min_{i\neq j}\lvert \tilde{E}_i-\epsilon_j\rvert \;\ge\; \tilde{\Delta}_j-\tilde{\varepsilon}_E,\\
\tilde{\Delta}_j:=\min\{\tilde{E}_{j}-\tilde{E}_{j-1},\,\tilde{E}_{j+1}-\tilde{E}_j\},
\end{aligned}
\end{equation}
so $\tilde{g}>0$ whenever $\tilde{\varepsilon}_E<\tilde{\Delta}_j$ and $\tilde{E}_j$ is the unique closest eigenvalue to $\epsilon_j$ if $\tilde{\varepsilon}_E<\tilde{\Delta}_j/2$.

Define the contraction factor
\begin{equation}
\begin{aligned}
\eta_{\mathrm{fsm}}
\;:=\;&
\frac{\max_{i\neq j}\,\lvert 1- C(\tilde{E}_i-\epsilon_j)^2\rvert}
     {\lvert 1- C(\tilde{E}_j-\epsilon_j)^2\rvert}
\\[4pt]
\le\;&
\frac{1- C \tilde{g}^2}{\lvert 1- C \tilde{\varepsilon}_E^2\rvert}
\;<\;1.
\end{aligned}
\end{equation}

After applying and normalizing $F_k(\mathcal{H})$, the ground-state–style overlap bound carries over to the target excited state:
\begin{equation}
\label{eq:QFSM_overlap}
p_k
\;:=\;
\frac{\lvert\langle \tilde{E}_j\vert F_k(\mathcal{H})\vert\psi^{(0)}\rangle\rvert^2}{\lVert F_k(\mathcal{H})\vert\psi^{(0)}\rangle\rVert_2^2}
\;\ge\;
\frac{1}{1+\frac{1-\lvert a_j\rvert^2}{\lvert a_j\rvert^2}\,\eta_{\mathrm{fsm}}^{\,2k}}.
\end{equation}
A lower bound on the postselection success probability mirrors the power iteration case:
\begin{equation}
\label{eq:QFSM_succ}
p^{(k)}_{\mathrm{succ}}
\;\ge\;
\frac{\bigl(\lvert a_j\rvert^2-(1-\lvert a_j\rvert^2)\,\eta_{\mathrm{fsm}}^{\,k}\bigr)^2}{\lvert a_j\rvert^2+(1-\lvert a_j\rvert^2)\,\eta_{\mathrm{fsm}}^{\,2k}}
\;\xrightarrow{k\to\infty}\; \lvert a_j\rvert^2.
\end{equation}

To reach overlap $1-\varepsilon$, it suffices that
\begin{equation}
\label{eq:QFSM_k_exact}
k \;\ge\; \frac{1}{2\log(1/\eta_{\mathrm{fsm}})}\;\log\!\frac{1}{\gamma^{2}\varepsilon}.
\end{equation}
With the conservative choice $C=1/\tilde{R}^2$ and in the regime $0 < C\tilde{\varepsilon}_E^2 < C \tilde{g}^2\,\ll 1$, we have $\log(1/\eta_{\mathrm{fsm}}) \approx  C\,(g^2-\varepsilon_E^2)$, yielding the sufficient estimate
\begin{equation}
\label{eq:QFSM_k_approx}
k \;\gtrsim\; \frac{\tilde{R}^{2}}{2\,(\tilde{g}^{2}-\tilde{\varepsilon}_E^{2})}\;\log\!\frac{1}{\gamma^{2}\varepsilon}.
\end{equation}

Implement $F_k(\mathcal{H})$ as a single QSVT polynomial of degree $d=2k$ in $\mathcal{H}=H/\lambda_\text{LCU}$.
As in previous sections, AA with $m_\star\simeq \pi/(4\gamma)-1/2$ boosts postselection, multiplying the polynomial cost by $(2m_\star+1)$.
The state-preparation query complexity is therefore
\begin{equation}
\label{eq:QFSM_queries_exact}
\text{Queries}_{\mathrm{QFSM+AA}}
\;=\;
\Theta\!\left(\frac{\pi}{4\gamma}\cdot d\right)
\;=\;
\Theta\!\left(\frac{\pi}{2\gamma}\cdot k\right),
\end{equation}
which, combined with \eqref{eq:QFSM_k_exact}, and returning back to the original units gives
\begin{equation}
\label{eq:QFSM_queries}
\begin{aligned}
\text{Queries}_{\mathrm{QFSM+AA}}
&\;=\;
\Theta\!\Biggl(
\frac{1}{\gamma}\cdot
\frac{\log\!\bigl(1/(\gamma^{2}\varepsilon)\bigr)}
{\log\!\bigl(1/\eta_{\mathrm{fsm}}\bigr)}
\Biggr)
\\[4pt]
&=
O\!\Biggl(
\frac{R^{2}}{\gamma\,(g^{2}-\varepsilon_E^{2})}\;
\log\!\frac{1}{\gamma^{2}\varepsilon}
\Biggr).
\end{aligned}
\end{equation}
In the regime, when $\varepsilon_E < \Delta_j/2$ we have $g=\Theta(\Delta_j)$ and $g^2 - \varepsilon_E^2 \geq (\Delta_j - 2\varepsilon_E)\Delta_j$.
Then using $R\lesssim 2\lambda_\text{LCU}$ one can get estimation:
\begin{equation}
\label{eq:QFSM_queries_final}
\begin{aligned}
\text{Queries}_{\mathrm{QFSM+AA}}
=
O\!\Biggl(
\frac{\lambda_\text{LCU}^{2}}{\gamma\,(\Delta_j - 2\varepsilon_E)\Delta_j}\;
\log\!\frac{1}{\gamma^{2}\varepsilon}
\Biggr).
\end{aligned}
\end{equation}

Unlike QPI, QPL, and QII, which target the ground state, QFSM prepares excited states using only an energy estimate $\epsilon_j$.
Its scaling in the spectral parameters is generally not optimal (it carries a $\lambda_\text{LCU}^{2}/((\Delta_j - 2\varepsilon_E)\Delta_j)$ factor rather than the first powers of gap and norm of the Hamiltonian), but it can be practical.
All query scalings and comparisons with existing algorithms are summarized in Table~\ref{tab:query-comparison}.

The query complexity table indicates that the GQSP–based filters proposed here are competitive with near-optimal ground state algorithms while remaining simple to implement and tune.
QPL+AA achieves ground state preparation with complexity $O({\lambda}/{(\Delta\,\gamma)} \log (1/\varepsilon))$ without any prior energy estimate or a costly energy search over shifts. 
This makes QPI a practical default when $E_0$ is unknown and one wants a guaranteed, monotone improvement in overlap at a cost that matches the standard $1/(\gamma\Delta)$ scaling up to constants.
Lanczos step reduces the iteration count by a constant factor $\beta>1$ relative to QPI while preserving the same asymptotic. 
In practice, this translates into tangible depth savings (tens of iterations in our numerics) with no extra oracles beyond those already required for QPI.

When a rough energy is available, inverse filtering wins.
If an a priori estimate satisfies $\varepsilon_E<\Delta/2$, the number of queries for single-shot inverse polynomial QII is estimated in Eq.~\eqref{eq:QIIpoly_queries_final}. 
As $\varepsilon_E$ improves, the effective gap $(\Delta-\epsilon_E)$ grows and the cost drops toward the $1/(\gamma\Delta)$ frontier.

QFSM extends the toolbox to excited states using the folded spectrum filter $(I-C(H-EI)^2)^k$. 
Although its dependence on spectral parameters is not optimal, it is operationally attractive. 
For moderate target fidelities, good $E$ (within the local gap), and re-use across multiple excitations, this simplicity can outweigh the non-optimal scaling.

The methods discussed above highlight distinct operating regimes. 
QPI and its Lanczos-enhanced variant QPL require no prior spectral information and provide reliable ground state preparation with modest depth. 
When an energy estimate with error $\varepsilon_E \leq \Delta/2$ is available, QII becomes the method of choice: its contraction factor improves rapidly as the estimate sharpens, and its overall query complexity approaches the near-optimal gap-inverse scaling.
For preparing specific excited states, QFSM offers a complementary strategy. 
Provided an energy guess lies inside the local spectral gap of the target level, its polynomial filter selectively amplifies the desired component. 
Together, these methods form a practical toolkit for state preparation across the range of situations encountered in Hamiltonian simulation and quantum chemistry.

\subsection{Resource estimation}
\label{subsec:resource_estimation}

Since the proposed methods are intended for execution in FTQC setting, we estimate the corresponding logical resource requirements. 
In particular, we report the T-gate count and the number of fault-tolerant (logical) qubits.

To estimate the total number of T gates, we decompose each Toffoli gate into $7$ T gates. 
The T-count for a single arbitrary rotation is estimated as
\begin{equation}
    T_R \approx 3 \log_2 \!\left(\frac{1}{\varepsilon_r}\right) + \mathrm{const},
\end{equation}
where $\varepsilon_r$ is the synthesis precision of the rotation. For the present order-of-magnitude estimates, the additive constant is neglected.

We allocate a uniform synthesis-error budget $\varepsilon_r$ across all arbitrary rotations. We take
\begin{equation}
    \varepsilon_r = \frac{\varepsilon_E}{\lambda_{\mathrm{LCU}} N_R},
\end{equation}
where $\lambda_{\mathrm{LCU}}$ is the Hamiltonian LCU norm, $N_R$ is the total number of rotational gates and $\varepsilon_E \approx 10^{-3}$ Ha is used as a reference chemical-accuracy scale. 
For the considered systems, the largest value is obtained for N$_2$, with $\lambda_{\mathrm{LCU}} = O(10^2)$.
This prescription should be regarded as a conservative resource-estimation proxy rather than a rigorous bound on the final energy error.

The resulting logical-resource estimates are summarized in Table~\ref{tab:resources}. 
T-counts should be interpreted primarily as estimates of the relative logical circuit cost of the proposed methods. 
Converting them into physical resources requires specifying both an error-correcting architecture and a physical error model.

All T-gate counts are reported per single iteration of the quantum power method. 
Therefore, the total T-count scales linearly with the number of iterations, while the dependence on the target precision enters logarithmically through the rotation-synthesis cost.
Moreover, these counts refer to a single coherent application of the
filter circuit. 
They do not include the repeated runs required by postselection, which increase the expected cost by a factor of order $1/p_{\mathrm{succ}}$, or
the additional circuit applications in amplitude amplification, which increase it
by a factor of order $1/\sqrt{p_{\mathrm{succ}}}$. 
They also exclude the overhead of the error-correcting layer itself, such as magic-state distillation, routing, and decoding, and the qubit counts in Table~\ref{tab:resources} include
only the logical qubits required by the algorithm.

\begin{table}[t]
\centering
\renewcommand{\arraystretch}{1.2}
\caption{Logical-resource estimates for one iteration of the quantum power method.
DZ denotes the cc-pVDZ basis set, and 6G denotes STO-6G.}
\label{tab:resources}
\begin{tabular}{llccc}
\hline\hline
Molecule & Basis & Active Space & T gates (1 iter) & Qubits \\
\hline\hline
H$_2$                    & DZ/$6$G & ($2$,$2$) & $1.3 \times 10^4$  & $9$  \\
CH$_2$                   & DZ      & ($6$,$6$) & $7.9 \times 10^5$ & $23$ \\
N$_2$                    & DZ      & ($6$,$6$) & $5.1 \times 10^5$ & $22$ \\
C$_2$H$_4$ ($0^\circ$)   & DZ      & ($2$,$2$) & $1.3 \times 10^4$  & $9$  \\
C$_2$H$_4$ ($90^\circ$)  & DZ      & ($2$,$2$) & $2.6 \times 10^4$  & $10$ \\
BeH$_2$                  & $6$G    & ($4$,$5$) & $1.9\times 10^5$ & $19$ \\
LiH                      & $6$G    & ($2$,$5$) & $3.7 \times 10^5$ & $20$ \\
H$_4$                    & $6$G    & ($4$,$4$) & $2.1 \times 10^5$ & $17$ \\
\hline\hline
\end{tabular}
\end{table}

The estimated T-count per iteration ranges from $O(10^4)$ to $O(10^6)$, while the number of logical qubits remains relatively modest ($9$--$23$) for the considered active spaces.
These values can be compared with the cost of QPE estimated under the same gate-synthesis assumptions.
For QPE with target energy precision $\varepsilon_E$, the required phase-register size is approximately
$\lceil \log_2(\lambda_{\mathrm{LCU}}/\varepsilon_E)\rceil$ (up to a small constant overhead), which yields about $20$ ancilla qubits for chemical accuracy in the present cases.
For CH$_2$ and N$_2$, the resulting QPE T-count is of the order $10^{12}$, with a total logical qubit requirement of approximately $40$, including the phase register.
Therefore, when quantum power methods are used for state preparation, their per-iteration cost is much lower than the cost of a full QPE calculation.

\subsection{Sensitivity to implementation errors}

The analysis above assumes an ideal logical implementation of the GQSP circuits. 
In a fault-tolerant setting, physical errors are suppressed by quantum error correction, but the resulting logical operations are still implemented with finite accuracy.
Relevant sources of error include imperfect block encodings, synthesis of the single-qubit rotations, and logical errors accumulated over repeated applications of the circuit. 
Here, we provide a worst-case perturbative analysis of how such errors affect the postselection probability and the accuracy of the prepared state.

Consider an ideal degree-$d$ GQSP circuit $V_d$ and an implemented circuit $\widetilde{V}_d$. 
The circuit contains $d$ applications of the signal operator and $d+1$ single-qubit rotations (see Fig.~\ref{fig:GQSP_cirsuit}). 
Suppose that each implemented block-encoding query differs from the ideal one by at most $\varepsilon_{\mathrm{BE}}$ in operator norm, and let the error of the $j$-th single-qubit rotation be bounded by $\varepsilon_j$. 
Applying the standard telescoping-product inequality to the product of unitary factors gives
\begin{equation}
\label{eq:circuit_error}
\left\|\widetilde{V}_d - V_d\right\|_2
\leq
d\,\varepsilon_{\mathrm{BE}}
+
\sum_{j=0}^{d}\varepsilon_j
\equiv
\varepsilon_{\mathrm{circ}}(d).
\end{equation}

If a common synthesis accuracy $\varepsilon_r$ is used for all rotations, this reduces to
\begin{equation}
\varepsilon_{\mathrm{circ}}(d)
\leq
d\,\varepsilon_{\mathrm{BE}}
+
(d+1)\varepsilon_r.
\end{equation}

This bound is deliberately conservative, since it assumes coherent worst-case accumulation and does not rely on cancellation between individual errors. 
Nevertheless, it shows that, at fixed logical accuracy per operation, the implementation error grows at most linearly with the polynomial degree.

Let $\Pi$ denote the projector onto the successful ancilla outcome and define the ideal and implemented unnormalized postselected states as
\begin{equation}
\lvert \chi_d\rangle
=
\Pi V_d \lvert \Phi_0\rangle,
\qquad
\lvert \widetilde{\chi}_d\rangle
=
\Pi \widetilde{V}_d \lvert \Phi_0\rangle,
\end{equation}
where $\lvert \Phi_0\rangle$ contains the input state and initialized ancillas. 
Since projection cannot increase the norm,
\begin{equation}
\left\|
\lvert \widetilde{\chi}_d\rangle
-
\lvert \chi_d\rangle
\right\|_2
\leq
\varepsilon_{\mathrm{circ}}(d).
\end{equation}

Writing
\begin{equation}
p_{\mathrm{succ}}^{(d)}
=
\||\chi_d\rangle\|_2^2,
\end{equation}
the corresponding perturbation of the postselection probability satisfies
\begin{equation}
\left|
\widetilde{p}_{\mathrm{succ}}^{(d)}
-
p_{\mathrm{succ}}^{(d)}
\right|
\leq
2\sqrt{p_{\mathrm{succ}}^{(d)}}\,
\varepsilon_{\mathrm{circ}}(d)
+
\varepsilon_{\mathrm{circ}}^2(d).
\end{equation}

Finite implementation errors therefore perturb the success probability continuously, but the normalized postselected state can be more sensitive when $p_{\mathrm{succ}}$ is small. 
Defining
\begin{equation}
\lvert \psi_d\rangle
=
\frac{\lvert \chi_d\rangle}
{\sqrt{p_{\mathrm{succ}}^{(d)}}},
\qquad
\lvert \widetilde{\psi}_d\rangle
=
\frac{\lvert \widetilde{\chi}_d\rangle}
{\sqrt{\widetilde{p}_{\mathrm{succ}}^{(d)}}},
\end{equation}
one obtains, provided
\begin{equation}
\varepsilon_{\mathrm{circ}}(d)
<
\sqrt{p_{\mathrm{succ}}^{(d)}},
\end{equation}
that
\begin{equation}
\left\|
\lvert \widetilde{\psi}_d\rangle
-
\lvert \psi_d\rangle
\right\|_2
\leq
\frac{
2\varepsilon_{\mathrm{circ}}(d)
}{
\sqrt{p_{\mathrm{succ}}^{(d)}}
}.
\end{equation}

Thus, the low postselection probabilities characteristic of nonunitary polynomial filters affect not only the sampling or amplitude amplification overhead but also their sensitivity to finite implementation errors.

Of particular relevance to the numerical results in this work is the resulting energy error. 
Let
\begin{equation}
E_d
=
\langle \psi_d \rvert H \lvert \psi_d\rangle,
\end{equation}
and define the energy variance of the ideal prepared state by
\begin{equation}
\sigma_d^2
=
\langle \psi_d \rvert
(H-E_d)^2
\lvert \psi_d\rangle.
\end{equation}
If
\begin{equation}
\label{eq:state_error}
\eta_d
=
\left\|
\lvert \widetilde{\psi}_d\rangle
-
\lvert \psi_d\rangle
\right\|_2,
\end{equation}
then normalization of both states gives
\begin{equation}
\label{eq:energy_error}
\left|
\widetilde{E}_d-E_d
\right|
\leq
2\sigma_d\eta_d
+
W\eta_d^2,
\end{equation}
where
\begin{equation}
W
=
E_{\max}-E_{\min}
\end{equation}
is the spectral width of $H$. 
Combining Eqs.~\ref{eq:state_error} and \ref{eq:energy_error} yields
\begin{equation}
\label{eq:energy_error_2}
\left|
\widetilde{E}_d-E_d
\right|
\leq
\frac{
4\sigma_d\,\varepsilon_{\mathrm{circ}}(d)
}{
\sqrt{p_{\mathrm{succ}}^{(d)}}
}
+
\frac{
4W\,\varepsilon_{\mathrm{circ}}^2(d)
}{
p_{\mathrm{succ}}^{(d)}
}.
\end{equation}

Equation~\ref{eq:energy_error_2} shows that state errors and energy errors behave differently close to convergence. 
As the polynomial filter approaches an eigenstate, $\sigma_d\rightarrow 0$, and the first-order contribution to the energy error vanishes. 
For an exactly prepared eigenstate, the energy error therefore becomes second order in the state-preparation error,
\begin{equation}
\left|
\widetilde{E}_d-E_d
\right|
=
O\!\left(
\frac{
W\,\varepsilon_{\mathrm{circ}}^2(d)
}{
p_{\mathrm{succ}}^{(d)}
}
\right).
\end{equation}
Consequently, a finite state-preparation error does not necessarily produce an energy error of the same order once the filtered state is already close to the target eigenstate.

This result also reveals a trade-off with polynomial degree.
Increasing $d$ improves the spectral selectivity of the polynomial filter, but simultaneously increases the number of imperfect logical operations. 
For QPI, for example, Eq.~\ref{eq:QPI_overlap} implies that the ideal ground-state infidelity $q_d$ obeys
\begin{equation}
q_d
\equiv
1-
\left|
\langle \widetilde{E}_0 \vert \psi_d\rangle
\right|^2
\leq
\frac{1-\gamma^2}{\gamma^2}
\rho^{2d}.
\end{equation}
Since the energy of any normalized state with ground-state infidelity $q$ satisfies
\begin{equation}
0
\leq
E_d-E_0
\leq
Wq_d,
\end{equation}
the energy of an imperfectly prepared QPI state can be bounded as
\begin{equation}
\label{eq:QPI_energy_error}
0
\leq
\widetilde{E}_d-E_0
\leq
W
\left[
\sqrt{
\frac{1-\gamma^2}{\gamma^2}
}
\rho^d
+
\frac{
2\varepsilon_{\mathrm{circ}}(d)
}{
\sqrt{p_{\mathrm{succ}}^{(d)}}
}
\right]^2.
\end{equation}

Hence, in the ideal calculation the error decreases exponentially with $d$, whereas for finite logical accuracy the accumulated implementation error eventually increases with circuit degree. 
If $p_{\mathrm{succ}}^{(d)}$ varies slowly and the rotation synthesis errors are approximately constant, Eq.~\ref{eq:QPI_energy_error} has the schematic form
\begin{equation}
\widetilde{E}_d-E_0
\lesssim
W
\left(
A\rho^d
+
Bd\varepsilon_\text{BE}
\right)^2.
\end{equation}

This implies that finite implementation accuracy generally introduces an optimal useful polynomial degree: beyond the point at which the ideal filtering error becomes smaller than the accumulated circuit error, increasing the degree no longer improves the final accuracy. 
The crossover occurs approximately when
\begin{equation}
\rho^d
\sim
d\varepsilon_\text{BE},
\end{equation}
up to factors associated with the initial overlap and postselection probability.

The same competition is present in the other methods considered here. 
For QPL, a degree-$k$ Lanczos polynomial followed by $n$ power steps has total polynomial degree approximately $d=n+k$. 
Increasing $k$ therefore introduces additional logical operations, but can substantially reduce the number of subsequent power iterations required to reach a target accuracy. 
The reduced total polynomial degree needed for a prescribed accuracy can therefore also improve robustness. 
For QII, the relevant circuit depth is controlled by the polynomial truncation degree used to approximate $(H-\epsilon_0 I)^{-n}$, rather than directly by the inverse-iteration order. 
Increasing the truncation degree reduces the polynomial-approximation error but increases $\varepsilon_{\mathrm{circ}}$, producing an analogous trade-off. 
For QFSM, the polynomial $[1-C(H-\epsilon_m I)^2]^n$ has degree $2n$. 
Quasi-degenerate folded eigenvalues require larger $n$ for spectral discrimination and are therefore also expected to be more sensitive to finite implementation accuracy.

Amplitude amplification introduces an additional accumulation of logical errors because the state-preparation circuit and its inverse are invoked repeatedly. 
If the underlying GQSP implementation has error $\varepsilon_{\mathrm{circ}}$, then a conservative worst-case bound after $m$ Grover iterations scales as
\begin{equation}
\varepsilon_{\mathrm{AA}}
=
O\!\left[
(2m+1)\varepsilon_{\mathrm{circ}}
\right],
\end{equation}
neglecting errors in the reflection operations themselves.
Since
\begin{equation}
m
=
O\!\left(
\frac{1}{\sqrt{p_{\mathrm{succ}}}}
\right),
\end{equation}
instances with very small success probability require correspondingly tighter logical gate accuracy. 
This reinforces the importance of improving both the initial overlap and the filter normalization, since these quantities influence resource cost and robustness simultaneously.

\subsection{Numerical comparison of QPI and QPL}
First, we performed numerical tests of the QPI and QPL methods. Because QPL can be viewed as a direct modification of QPI, we discuss both in this section. Following the notation introduced in Eq.~\eqref{eq:qpl4}, we denote the QPL method of the Lanczos polynomial degree $k$ by $\mathrm{QPL}k$. In this notation, the QPI method corresponds to $\mathrm{QPL}0$.

\begin{figure*}[ht] 
\includegraphics[width=1\textwidth]{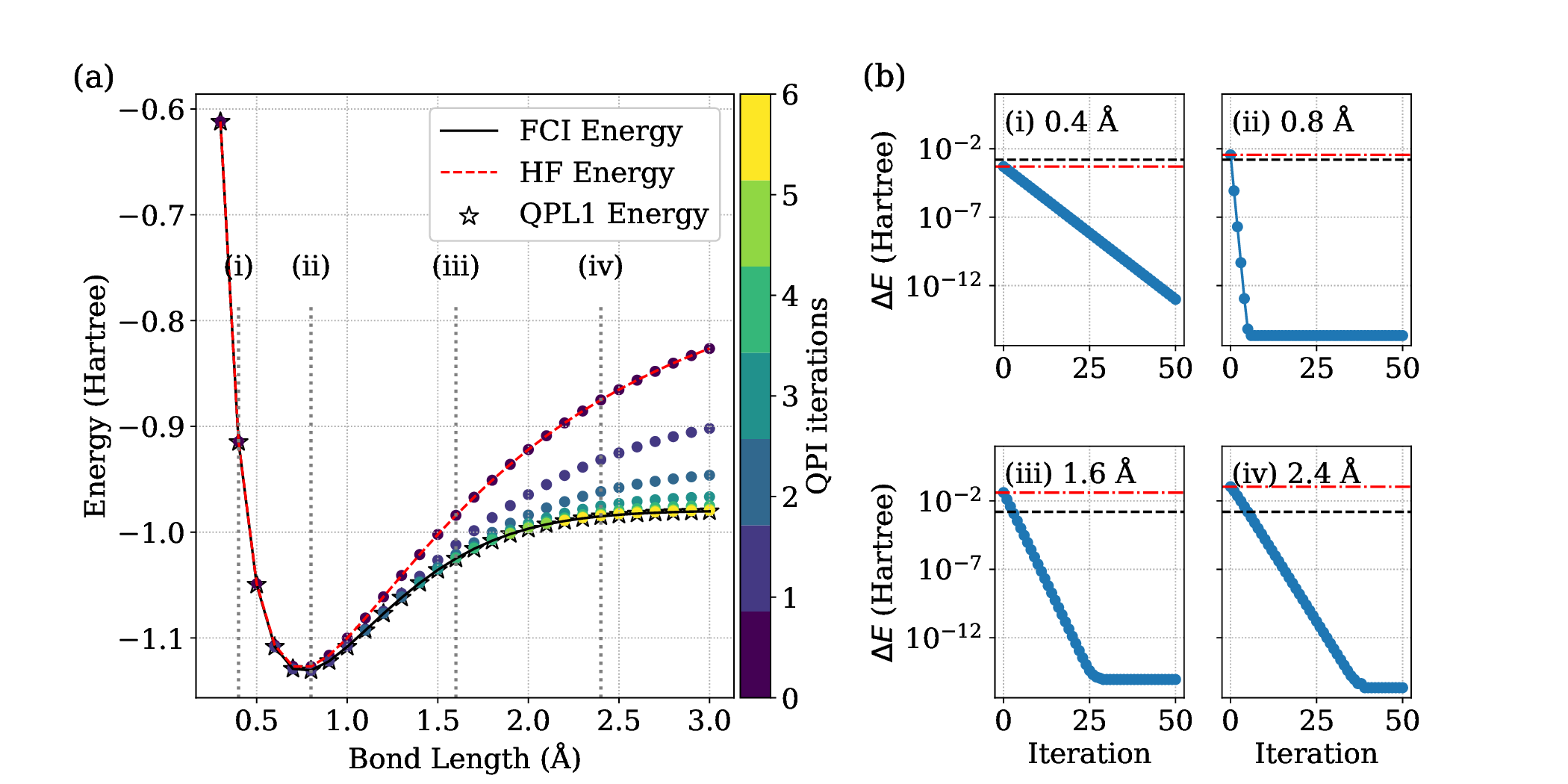} 
\caption{(a) Potential energy curve for the hydrogen molecule obtained with the QPI method ($\mathrm{QPL}0$). The basis set is cc-pVDZ, and the active space is $(2,2)$. 
Star markers indicate the corresponding $\mathrm{QPL}1$ energies immediately after the variational optimization step.
(b) The right panels show the energy deviation from CASCI$(2,2)$ at each iteration for the first $50$ iterations. The black dashed line indicates chemical accuracy ($1\,\text{kcal/mol}$), and the red dot-dashed line is the HF energy.} 
\label{fig:H2_QPL} 
\end{figure*}

Figure~\ref{fig:H2_QPL} illustrates the potential energy curves (PECs) for the hydrogen molecule using the cc-pVDZ basis set with $2$ electrons in $2$ active orbitals. The Hartree-Fock (HF) wave function served as the initial guess.

We observe that the QPI ($\mathrm{QPL}0$) method achieves the CASCI$(2,2)$ energy within chemical accuracy in no more than $6$ iterations, even when starting from the HF wave function. For longer bond lengths, more iterations are required to achieve chemical accuracy, which reflects the increased multireference character of the wavefunction. This behavior became more pronounced as the bond was stretched, indicating that the method may converge more slowly in systems with stronger static correlation.

The convergence speed of the power method depends on the ratio $E_1/E_0$, where $E_0$ and $E_1$ denote the ground state and first excited state energies, respectively. The panels on the right side of Fig.~\ref{fig:H2_QPL} show that the convergence rates are approximately proportional to $\log_{10}(E_1/E_0)$, with a proportionality coefficient between $1.85$ and $2$ for different bond lengths. 
This behavior is fully consistent with the classical power iteration algorithm, whose convergence is governed by the same ratio between largest eigenvalues. The quantum implementation therefore reproduces the expected classical scaling.

Next, we examine $\mathrm{QPL}1$, where the final wave function after optimizing the Lanczos polynomial at degree $1$ reads
\begin{equation}
\label{eq:H2_QPL}
    |\mathrm{QPL}1\rangle = \frac{1}{N} (a_0  + a_1 \mathcal{H}) |\mathrm{HF}\rangle.
\end{equation}
where $N$ is a normalization factor. Remarkably, this ansatz can match the CASCI energy immediately after the polynomial optimization and before any additional power method steps.
This means that QPL$1$ can reduce the number of block encoding queries. 
For example, at a bond length of $3$ \AA, QPI requires $6$ queries, whereas QPL$1$ achieves the same result with only one.

\begin{figure}[ht] 
\includegraphics[width=1\textwidth]{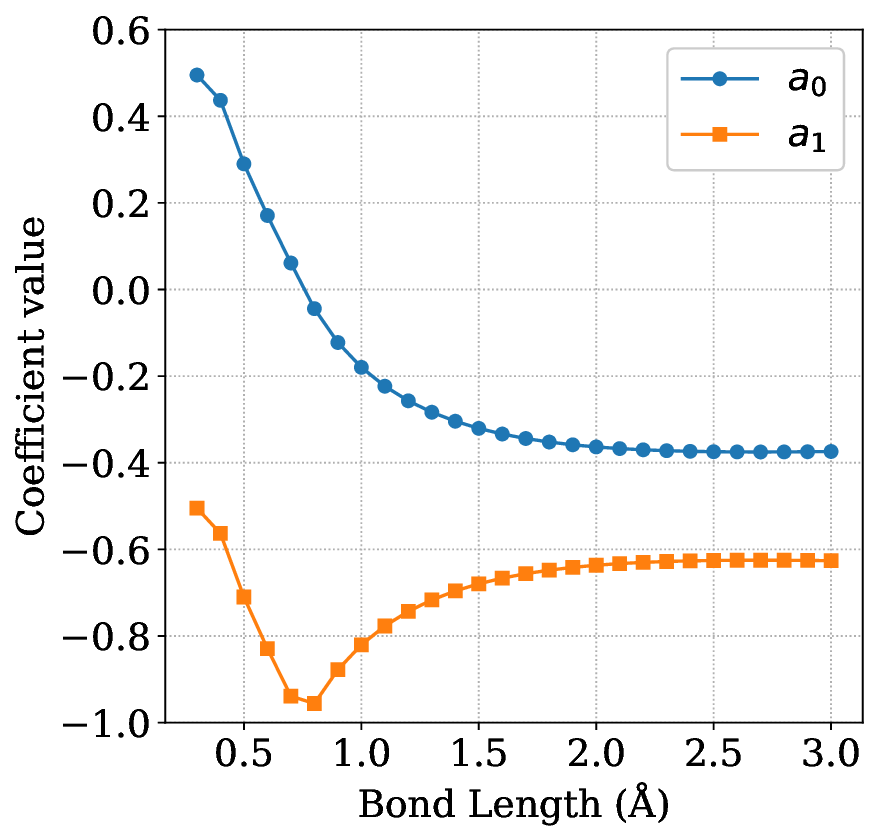} 
\caption{Bond length dependence of the Lanczos polynomial coefficients for the hydrogen molecule. Coefficients are obtained after the variational optimization of the Lanczos polynomial (see Eq.~\eqref{eq:H2_QPL}).} 
\label{fig:H2_QPL_coeffs} 
\end{figure}

Figure~\ref{fig:H2_QPL_coeffs} illustrates how the Lanczos polynomial coefficients $a_0$ and $a_1$ vary with bond length. We observe that $a_0$ changes sign around equilibrium geometry of H$_2$. Because $a_0$ represents the HF component of the wave function, its sign change reflects a decreasing HF contribution as the bond stretches. Meanwhile, $a_1$, which weighs $\mathcal{H}|\mathrm{HF}\rangle$, reaches its maximum near the equilibrium geometry, and then decreases. This behavior is expected because $\mathcal{H}|\mathrm{HF}\rangle$ is not orthogonal to $|\mathrm{HF}\rangle$; therefore, the relative amplitudes of these two contributions can offset each other.

\begin{figure*}[ht] 
\includegraphics[width=1\textwidth]{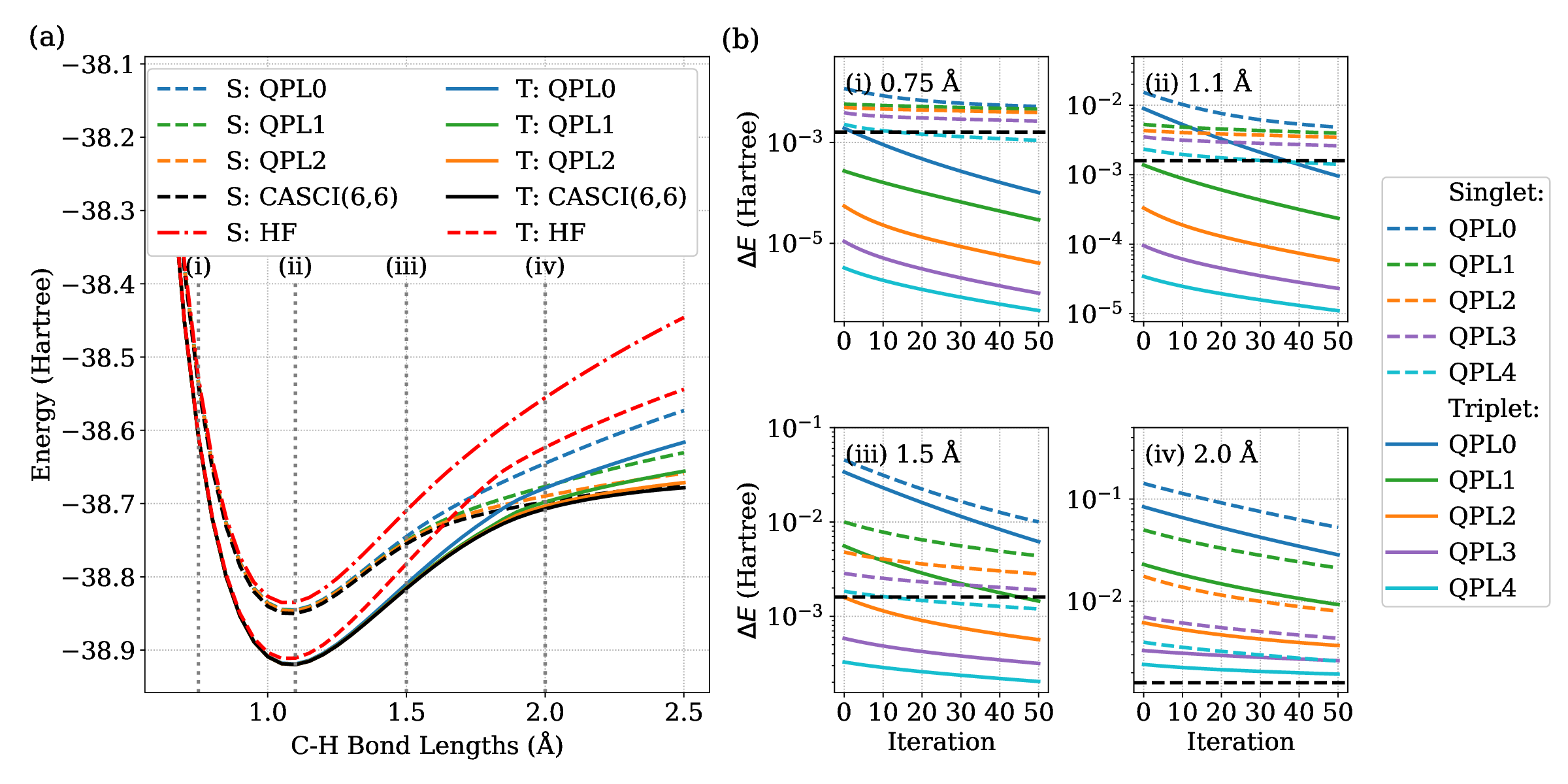} \caption{(a) Potential energy curves for singlet (S) and triplet (T) CH$_2$ obtained using QPL at various polynomial degrees ($k=0,1,2$). 
All $\mathrm{QPL}k$ curves correspond to energies after $50$ iterations.
The basis set is cc-pVDZ with an active space of $(6,6)$. 
(b) The right panels show deviations from CASCI$(6,6)$ for the first $50$ iterations of the power method using QPL with $k=0,1,2,3,4$. The dashed black lines indicate chemical accuracy ($1\,\text{kcal/mol}$). Solid lines correspond to the triplet state and dashed lines to the singlet state.} 
\label{fig:CH2_QPL} 
\end{figure*}

Subsequently, we computed the PECs of the singlet and triplet states of the CH$_2$ radical. A cc-pVDZ basis set with $6$ electrons in $6$ active orbitals was employed, and both C--H bonds were stretched symmetrically while keeping the H--C--H angle fixed at $101.89^\circ$. For the singlet state, we used the restricted Hartree-Fock (RHF) wave function as an initial guess, and for the triplet state, we used the restricted open-shell Hartree-Fock (ROHF) wave function. The results are presented in Fig.~\ref{fig:CH2_QPL}.

Notably, the triplet state lies below the singlet state around the equilibrium geometry of CH$_2$. For both spin states, $\mathrm{QPL}k$ yielded a marked improvement over the respective HF initial guesses. However, unlike the hydrogen molecule, $\mathrm{QPL}0$ (QPI) failed to achieve chemical accuracy after $50$ iterations for bond lengths beyond approximately $1.5$ \AA\ in the triplet and for all bond lengths in the singlet. Hence, we need to use Lanczos polynomials to reach chemical accuracy. As the bond stretches, this requirement becomes more pronounced, which is consistent with the growing multireference character of the wave function.

There is a consistent advantage from using higher-degree Lanczos polynomials. 
In our simulations, each additional unit of polynomial degree typically replaces up to $50$ power iteration steps, in line with the constant-factor improvement predicted for QPL over QPI. 
In practice, however, optimizing polynomial coefficients becomes increasingly difficult at large $k$. 
For this reason, moderate polynomial degrees are generally preferable: pushing QPL to very high degrees risks turning it into a VQE-like procedure with hundreds of tunable parameters and diminishing practical benefit.
This observation is consistent with those reported in Ref.\cite{Power_Lanczos}. 

Finally, the singlet state, with its lower spin, displays a stronger multireference character than the triplet state, generally requiring higher degree Lanczos polynomials to achieve comparable accuracy.

\begin{figure*}[ht] 
\includegraphics[width=1\textwidth]{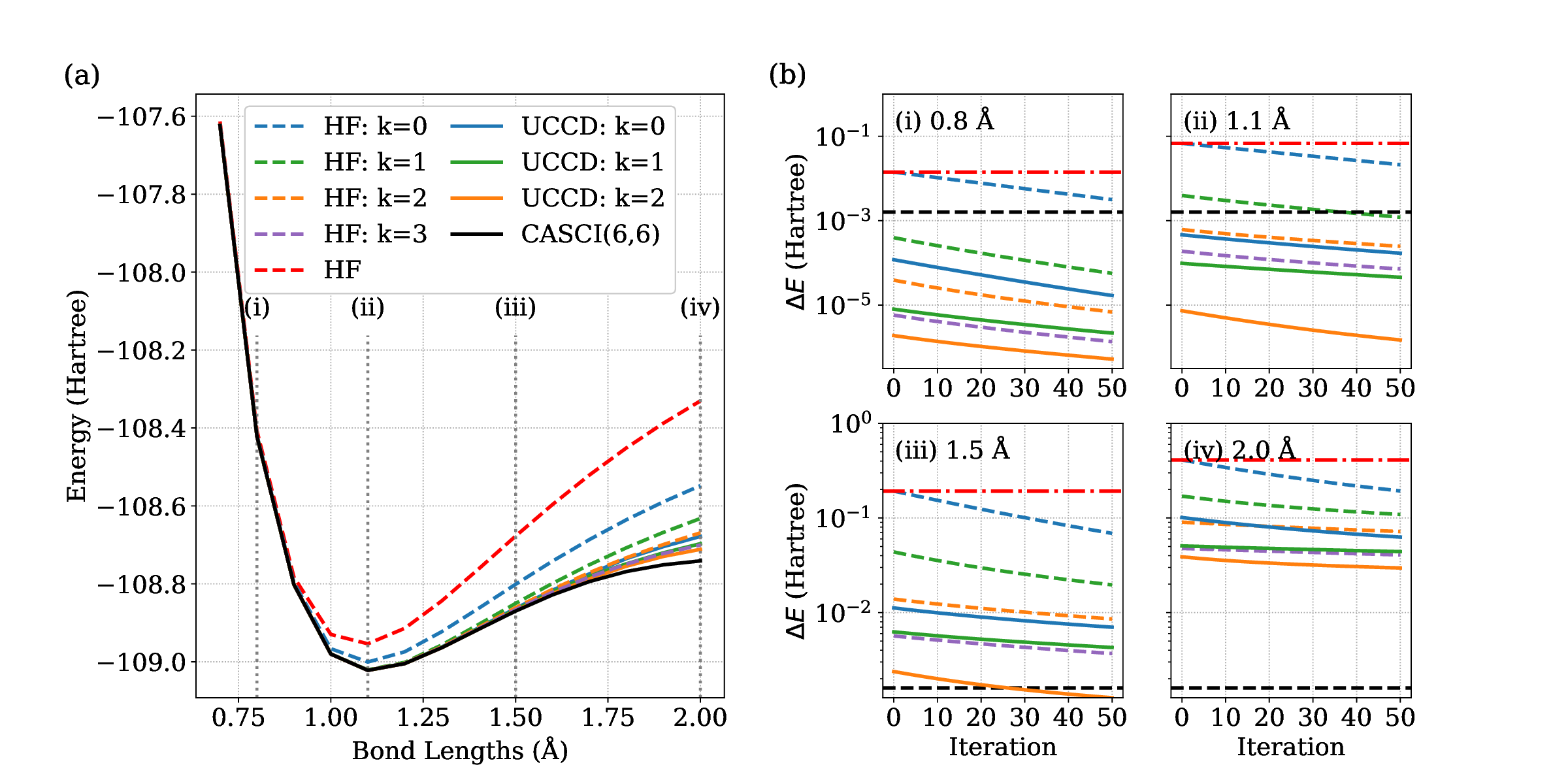} \caption{
(a) Potential energy curves for the nitrogen molecule obtained using QPL with different polynomial degrees ($k=0,1,2,3$) and initial guesses (HF in dashed, and partially converged UCCD in solid). 
All $\mathrm{QPL}k$ curves correspond to energies after $50$ iterations.
The basis set is cc-pVDZ, and the active space is $(6,6)$. 
(b) The right panels display energy differences from CASCI$(6,6)$ over the first $50$ iterations of the power method. The black dashed line marks chemical accuracy ($1\,\text{kcal/mol}$), and the red dot-dashed line is the HF energy.} 
\label{fig:N2_QPL} 
\end{figure*}

As a further test, we applied QPL to a nitrogen molecule, which is known to have a highly multireference ground state character in stretched geometries. The cc-pVDZ basis set with $6$ electrons in $6$ active orbitals was used. Two different initial guesses were tested:
\begin{enumerate} 
\item The RHF wave function. 
\item A partially converged VQE wave function with a unitary coupled cluster method with double excitations only (UCCD) ansatz (after only $2$ iterations), which is thus multireference but not fully optimized. 
\end{enumerate} 
The VQE optimization was halted after two optimization steps, without reaching a convergence threshold, so the resulting state is an intermediate, not fully optimized, multireference wave function.
The corresponding PECs are presented in Fig.~\ref{fig:N2_QPL}.

For the equilibrium geometry, $\mathrm{QPL}0/\mathrm{HF}$ cannot match the CASCI$(6,6)$ energy within $50$ iterations, but $\mathrm{QPL}1/\mathrm{HF}$ succeeds. However, even at longer bond lengths (1.5 \AA\ and 2.0 \AA), $\mathrm{QPL}3/\mathrm{HF}$ does not achieve chemical accuracy. These results reiterate that highly multireference systems pose convergence challenges for QPI and QPL, although higher-degree polynomials are often helpful.
However, each Lanczos polynomial of degree $k$ achieves a better result after a single iteration than a polynomial of degree $k-1$ does after $50$ iterations.
This reflects the constant advantage provided by QPL.

A more effective approach is to use a better initial guess than HF. Indeed, starting from the partially converged UCCD wave function significantly improves the convergence, allowing us to achieve chemical accuracy at longer bond lengths. However, for stretched geometries (for example, $2.0$ \AA\ in Fig.~\ref{fig:N2_QPL}), although this improved initial guess is insufficient without further refinement.
The constant advantage offered by QPL is evident here as well. 
At a bond length of $1.1$ \AA, where $\mathrm{QPL}2/\mathrm{UCCD}$ after a single iteration yields an error nearly an order of magnitude lower than $\mathrm{QPL}1/\mathrm{UCCD}$ after $50$ iterations.

In addition, these results demonstrate that QPI and QPL are far more effective in refining a reasonably good reference wave function that captures most of the static correlation rather than attempting to recover both static and dynamic correlation from a simple HF state. This use case motivated for the original power Lanczos approach in Ref.~\cite{Power_Lanczos}, in which the method was used to add a dynamic correlation to Monte Carlo wave functions that already contain strong static correlation.

\subsection{Numerical Results for QII}
In this section,the performance of the GQSP-based QII algorithm is described. First, we present the PEC calculation as an example. We then compared our QII implementation with two existing methods, I-Iter and Q-Inv, which rely on linear combinations of imaginary-time evolution operators \cite{Cainelli2024}.

\begin{figure*}[ht] 
\includegraphics[width=1\textwidth]{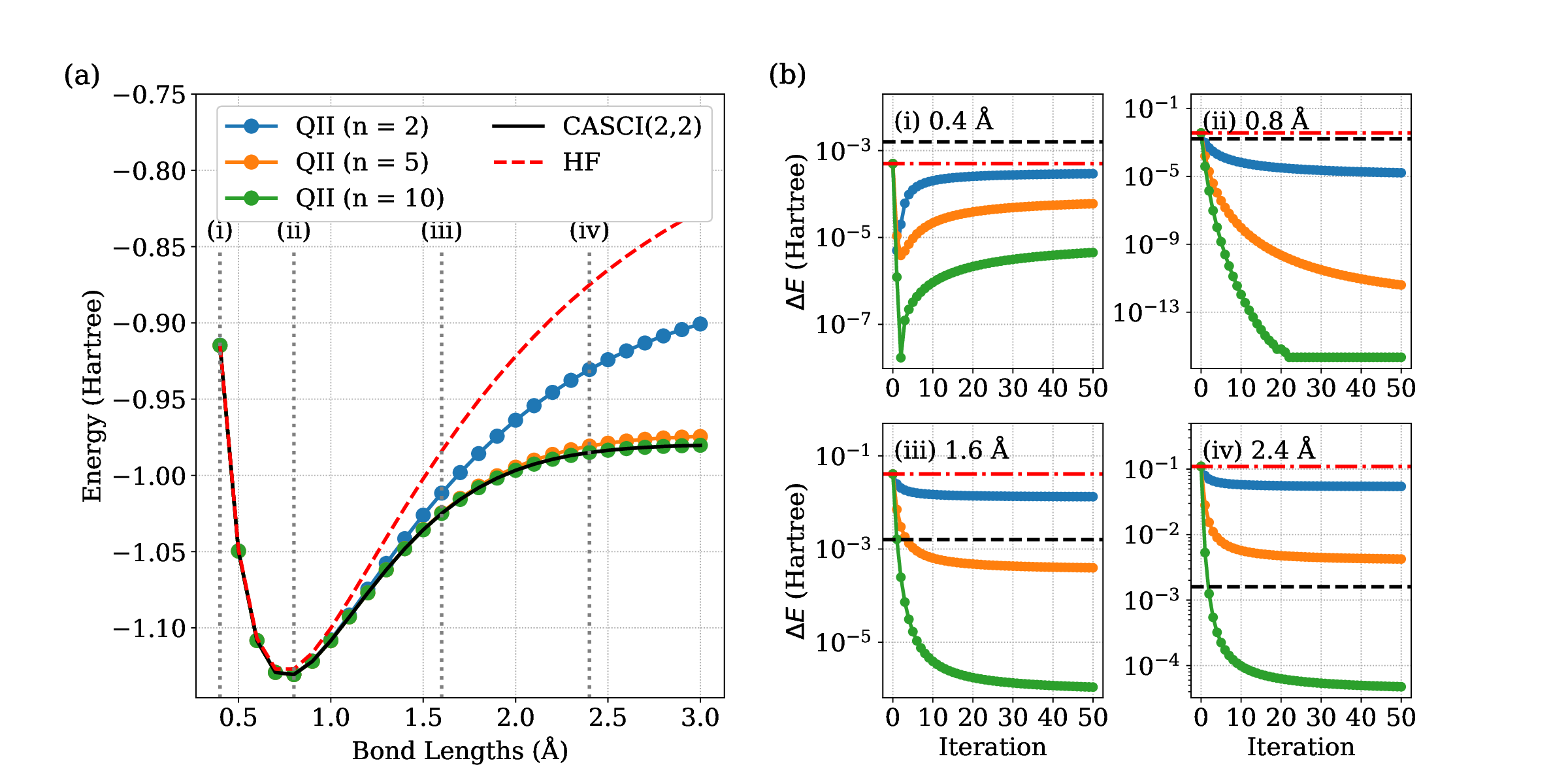} \caption{
(a) Potential energy curves for the hydrogen molecule obtained using QII with different polynomial truncation degrees ($n=2,5,10$). 
The HF energy was used as an energy shift.
All QII curves correspond to energies after $50$ iterations.
The basis set is cc-pVDZ, and the active space is $(2,2)$. 
(b) The right panels show the deviation from the CASCI$(2,2)$ energy over the first 50 iterations. The black dashed lines mark chemical accuracy ($1\,\text{kcal/mol}$), and the red dot-dashed lines represent the HF energy.} 
\label{fig:H2_QII} 
\end{figure*}

We begin by examining the dissociation curve of the hydrogen molecule. The calculations used the cc-pVDZ basis set with $2$ electrons in $2$ active orbitals. In QII, the inverse operator is approximated by truncating the Taylor polynomial to a certain degree $n$. The results for $n=2$, $n=5$, and $n=10$ are presented. 
It is important to note that the number of block encoding queries equals the polynomial degree $n$ and does not depend on the iteration count.
The number of iterations only determines the shape of the approximated function.
QII also requires an initial estimate of the ground state energy, which we consider to be the HF energy in all calculations. 
Figure~\ref{fig:H2_QII} shows the resulting PECs.

We observed that the ability to reach the CASCI$(2,2)$ energy within chemical accuracy depends on the degree $n$ of the truncated polynomial. For instance, at a bond length of $1.6$ \AA, $\mathrm{QII}(n=2)$ converges to a value above the chemical accuracy; similarly, $\mathrm{QII}(n=5)$ struggles at $2.4$ \AA. Conversely, for a sufficiently large $n$ ($n=50$), the method achieves chemical accuracy in only one iteration for all bond lengths. 
This rapid convergence relative to the QPI method reflects the well-known classical advantage of inverse iteration when a good approximation for the ground state energy estimation is available.
In other words, the faster convergence is not specific to the quantum adaptation but follows the traditional hierarchy between inverse and power iterations.

Next, we compare our GQSP implementation of QII with the I-Iter and Q-Inv algorithms, which evaluate $H^{-1}$ or $H^{-k}$ by numerically integrating the imaginary-time evolution operators \cite{Cainelli2024}. The I-Iter method approximates $H^{-1}$ using
\begin{equation}
\label{eq:I-Iter}
    {H}^{-1} = \frac{i}{\sqrt{2\pi}} \int_0^\infty \mathrm{d}y \int_{-\infty}^{\infty} \mathrm{d}z\, z\, e^{-z^2/2} e^{-i y z {H}}.
\end{equation}
and repeated application of $H^{-1}$ ($k$ times) yields $H^{-k}$. The Q-Inv method proposed by Kyriienko \cite{Kyriienko2020} directly approximates $H^{-k}$ by
\begin{equation}
\label{eq:Q-Inv}
    {H}^{-k} = \frac{i N_k}{\sqrt{2\pi}} \int_0^\infty \mathrm{d}y \int_{-\infty}^{\infty} \mathrm{d}z\, z\, y^{k-1} e^{-z^2/2} e^{-i y z {H}},
\end{equation}
where $N_k$ is the normalization constant. 
Both I-Iter and Q-Inv use the Gauss-Legendre rule to calculate the integrals, and the precision of this procedure depends on the order of the Legendre polynomials used. 
In Ref.~\cite{Cainelli2024}, comparisons were made for two orders $n_y$ for the integration of the variable $y$, in particular $n_y=1$ and $n_y=n_{\mathrm{eq}}$, where $n_{\mathrm{eq}}$ is the order sufficient to achieve energy convergence.
The hybrid Q-Inv+I-Iter strategy was also introduced. 
In this approach, the Q-Inv algorithm is first applied up to the iteration order $k$ at which its numerical error reaches a minimum, since the accuracy of the direct $H^{-k}$ approximation deteriorates for larger $k$ due to limitations in the integral discretization. 
Once this optimal point is reached, a small number of subsequent I-Iter steps are performed to further suppress the remaining error. 
This hybrid procedure leverages the rapid early-stage convergence of Q-Inv while avoiding its high-$k$ breakdown, and it reduces the total cost of inverse iterations because only a few applications of the I-Iter update are required.

\begin{figure*}[ht] 
\includegraphics[width=1\textwidth]{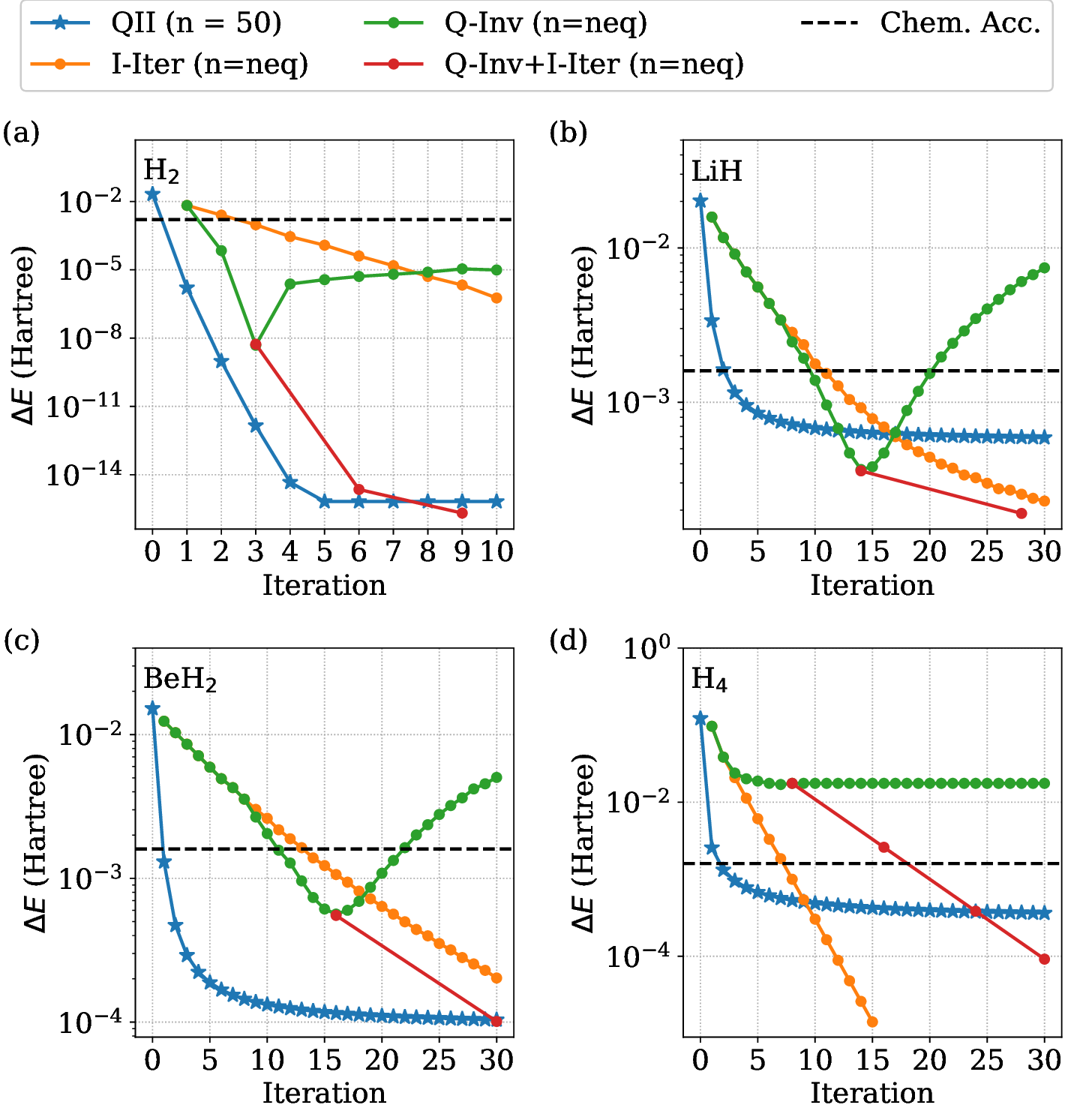} \caption{Comparison of QII, I-Iter, and Q-Inv \cite{Cainelli2024} for H$_2$, LiH, BeH$_2$, and H$_4$. The deviation from CASCI energies as a function of the iteration number is shown. For QII, $n=50$ refers to the Taylor polynomial truncation degree, while for I-Iter and Q-Inv, $n_y$ is the Gauss-Legendre polynomial order. $n_{\mathrm{eq}}$ is the order at which energy convergence is reached.} 
\label{fig:Comparison_QII_short} 
\end{figure*}

Analogous to the reference paper, we investigated four molecules: H$_2$, LiH, BeH$_2$, and a square H$_4$, using a STO-$6$G basis set. 
For H$_2$ and H$_4$ full active spaces were used $(2,2)$ and $(4,4)$, respectively. For LiH, $2$ electrons were used in $5$ orbitals, while $4$ electrons were used in $5$ orbitals for BeH$_2$. 
All calculations were single-point with geometries of $0.75$ \AA\ for H$_2$, $1.23$ \AA\ for the H--H distance in H$_4$, $1.6$ \AA\ for the Li--H bond in LiH, and $1.326$ \AA\ for the Be--H bond in BeH$_2$. 
Figure~\ref{fig:Comparison_QII_short} compares only QII with truncation $n=50$ with Q-Inv and I-Iter with $n_y=n_{\mathrm{eq}}$. The results for different truncations are discussed in Appendix~\ref{appendix:qii_truncation}. 
In the figure, the iteration number used for Q-Inv, I-Iter, and the hybrid Q-Inv+I-Iter refers to the power $k$ in the approximate application of $H^{-k}$. 
For I-Iter, this corresponds to applying the numerically integrated approximation of $H^{-1}$ repeatedly $k$ times, whereas Q-Inv constructs a direct approximation of $H^{-k}$ through its Fourier-integral representation.
For QII, the iteration number only reflects the shape of the approximated function, while the number of queries remains constant at $50$.
When comparing the number of qubits between methods, we assumed LCU block encoding. 
We also include qubit counts using qubit tapering.

For the simplest system H$_2$, chemical accuracy was achieved quickly using all methods. QII and Q-Inv+I-Iter achieved machine precision. Q-Inv began to diverge after $3$ iterations. For QII the number of qubits is constant for all iterations here and is equal to $7$ (with qubit tapering $4$). The number of qubits for Q-Inv and I-Iter is equal to at least $13$. However, because we used a truncation degree $50$, we required the same number of queries as the block encoded Hamiltonian, which makes the circuit deep. Q-Inv and I-Iter do not require queries for block encoding; however, they require a Trotter approximation of the time-evolution operator, which can also be relatively deep.

For LiH, all methods eventually achieved chemical accuracy. However, QII performed this task in $3$ iterations, whereas Q-Inv and I-Iter required more than $10$. Q-Inv also began to diverge significantly after $14$ iterations. This divergent behavior was fixed in the combined implementation of Q-Inv and I-Iter. For QII the number of qubits was $11$ (with qubit tapering $7$). The number of qubits for Q-Inv and I-Iter was equal to at least $17$. We used the same truncation degree $50$; therefore, we required the same number of queries as the block encoded Hamiltonian.

For BeH$_2$,  QII achieved chemical accuracy in one iteration, whereas Q-Inv and I-Iter required more than $10$. Generally, Q-inv and I-Iter demonstrated a similar behavior to that of LiH. The resource requirements were also the same as those for LiH. QII requires $11$ qubits (with qubit tapering $7$), whereas Q-Inv and I-Iter require at least $17$. The number of queries for block encoding was the same as before.

The final tested system was the square H$_4$. It was a challenging case for Q-Inv since it converged before chemical accuracy was reached. QII and I-Iter were more successful. Although QII with $n=50$ demonstrates a very fast reach to the chemical accuracy with smaller truncation, it has the problem of early convergence, analogous to Q-Inv (see Appendix~\ref{appendix:qii_truncation}). QII requires $9$ qubits (with qubit tapering $7$), whereas Q-Inv and I-Iter require at least $15$. 

In summary, the proposed GQSP-based QII implementation offers several advantages. First, implementing the time-evolution operator in I-Iter and Q-Inv typically involves Suzuki--Trotter decompositions, which introduce additional errors and increase the circuit depth. Second, I-Iter and Q-Inv require LCU approaches to approximate the integral, necessitating $O(\log (n_y n_z))$ auxiliary qubits, where $n_y$ and $n_z$ correspond to the number of nodes in the numerical integration of Eq.~\eqref{eq:I-Iter} and Eq.~\eqref{eq:Q-Inv}. The authors noted that $n_z$ depends on the power of the inverse Hamiltonian $k$. Moreover, the application of $H^{-1}$ $k$ times using I-Iter is not explicitly explained. It may be necessary to perform state tomography after each step or the use of QSP. The same problem applies to Q-Inv+I-Iter. Conversely, GQSP requires only one signal processing qubit, one qubit for capitalization (if needed), and additionally $O(\log L)$ qubits for LCU block encoding, where $L$ is the number of Pauli terms in the LCU expansion (see Eq.~\eqref{eq:lcu1}). In the worst case, $L$ scales as $O(N_{\text{orb}}^4)$, leading to $O(\log N_{\text{orb}})$ ancilla qubits, where $N_{\text{orb}}$ denotes the number of orbitals in active space. Third, QII reaches chemical accuracy quickly, having provided a sufficiently large truncation value. However, QII requires a good initial estimate of the ground state energy, whereas I-Iter and Q-Inv do not. Nevertheless, if such an estimate is available, QII can converge to a high accuracy in very few iterations, particularly for larger polynomial degrees.

\subsection{Excited states calculations using QFSM}
Finally, we performed numerical tests of the Quantum Folded Spectrum Method (QFSM) implemented using GQSP. Unlike the other algorithms discussed in this study, QFSM enables access to excited states.

In the first test, we computed the PECs of the ground and excited states in the hydrogen molecule using the cc-pVDZ basis set with an active space of $2$ electrons in $2$ orbitals. The convergence of QFSM critically depends on the choice of the initial guess wave function, which should have a large overlap with the target excited state. For instance, starting from an RHF wave function does not allow convergence to the $1^3\Sigma_u^+$ triplet state or $1^1\Sigma_u^+$ singlet excited state.

\begin{figure*}[ht] 
\includegraphics[width=1\textwidth]{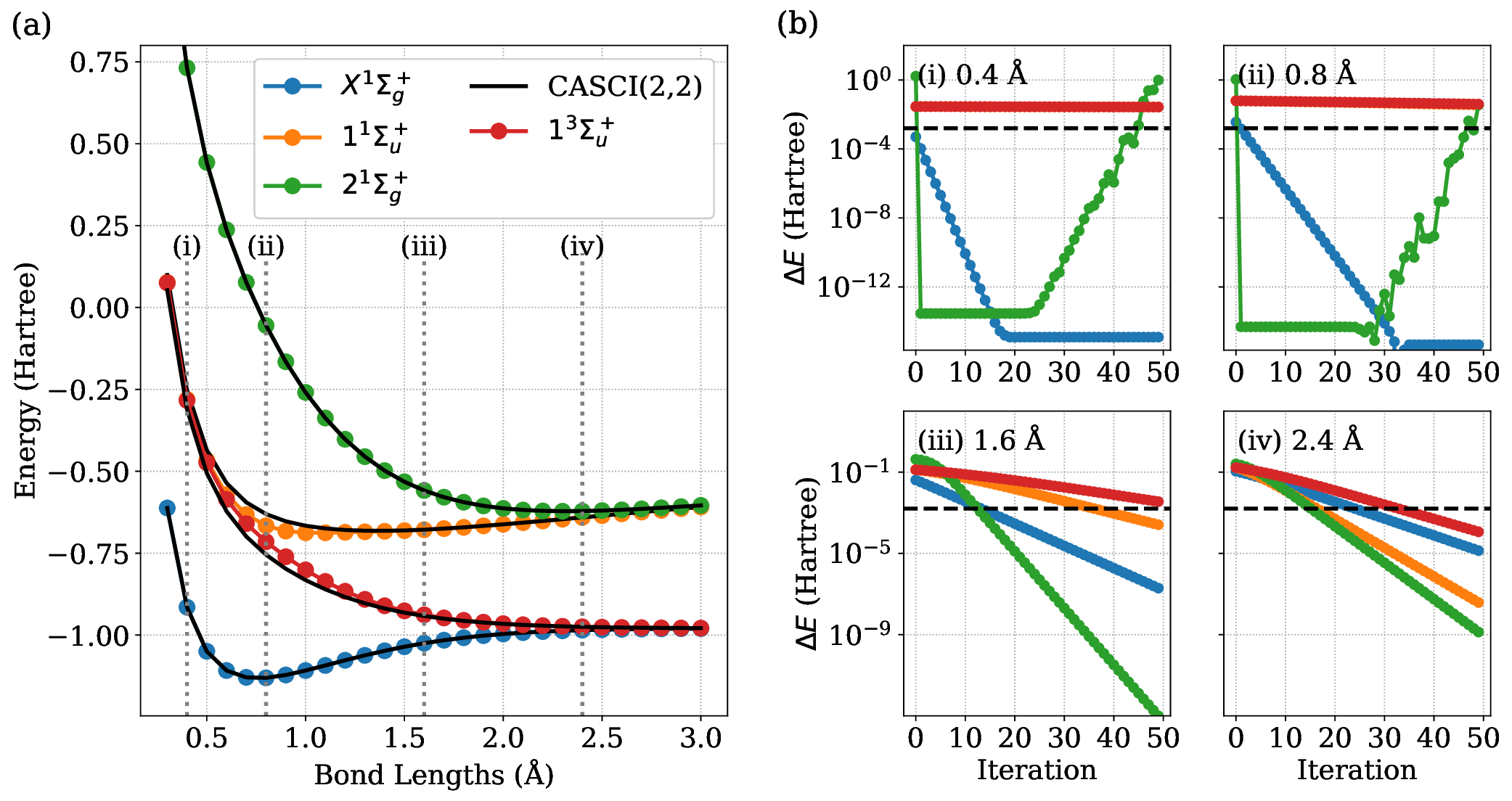} \caption{
(a) Potential energy curves for the hydrogen molecule computed using QFSM. The basis set is cc-pVDZ and the active space is $(2,2)$. 
All QFSM curves correspond to converged energies or energies after $50$ iterations, if convergence was not achieved.
(b) The right plots show the energy deviation from CASCI$(2,2)$ over the first 50 iterations. The black dashed line denotes chemical accuracy ($1\,\text{kcal/mol}$).} 
\label{fig:H2_QFSM} 
\end{figure*}

To ensure sufficient overlap with specific target states, we employed the following initial guess states:
\begin{equation}
\label{eq:state_QFMS_H2}
\begin{aligned}
    |\Psi_{\Sigma_g} \rangle = |0011\rangle, \\
    |\Psi_{\Sigma_u} \rangle = |1001\rangle ,
\end{aligned}
\end{equation}
where the ordering corresponds to $|\sigma_\beta^*\sigma_\alpha^*\sigma_\beta \sigma_\alpha \rangle$. As an energy shift in the folded Hamiltonian, we used the corresponding CASCI$(2,2)$ energies for testing. The results are shown in Fig.~\ref{fig:H2_QFSM}.

The singlet states $X^1\Sigma_g^+$ and $2^1\Sigma_g^+$ converged to chemical accuracy across all bond lengths when starting from $|\Psi_{\Sigma_g} \rangle$.
However, at $0.4$ \AA\ and $0.8$ \AA, divergence was observed after $25$ iterations owing to the loss of orthogonality in floating point arithmetic. This is a typical drawback of power iteration methods. 
Convergence to the $1^1\Sigma_u^+$ and $1^3\Sigma_u^+$ states was more challenging when starting from $|\Psi_{\Sigma_u} \rangle$. At short bond lengths ($0.4$ \AA\ and $0.8$ \AA), these states were unable to achieve chemical accuracy, and their convergence was very slow. The right-hand plots show overlapping data points for these two states, indicating near-degeneracy. This behavior can be attributed to the presence of quasi-degenerate eigenstates in the folded Hamiltonian, which causes the wave function to be simultaneously attracted toward both, thereby hindering the convergence. Improved initial guesses could resolve this issue.

\begin{figure*}[ht] 
\includegraphics[width=1\textwidth]{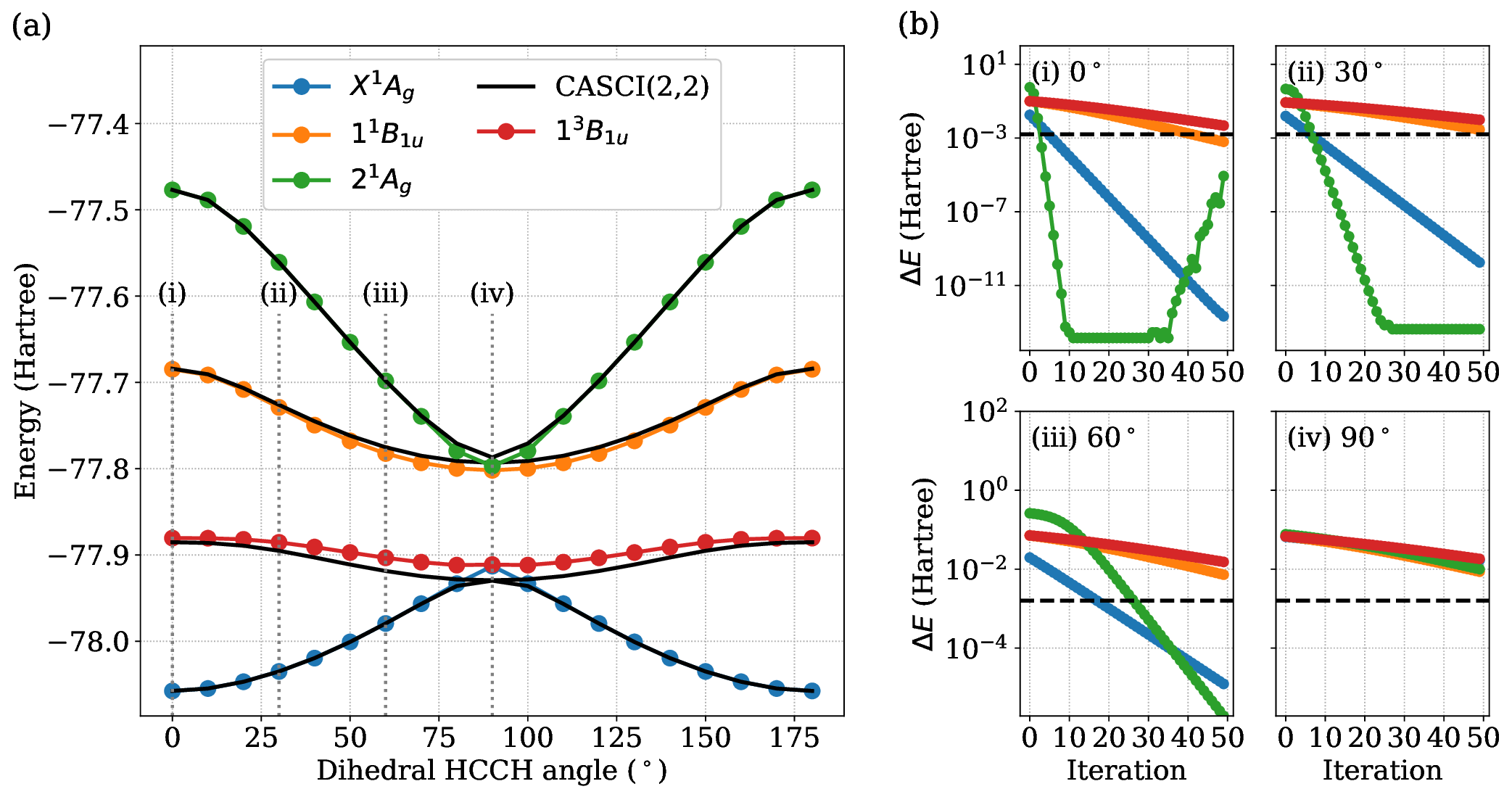} \caption{(a) Potential energy curves for ethylene along the torsional rotation of the double bond, computed using QFSM. The basis set is cc-pVDZ and the active space is $(2,2)$. 
(b) The left plots show energy deviation from CASCI$(2,2)$ over the first $50$ iterations. The black dashed line indicates chemical accuracy ($1\,\text{kcal/mol}$).} 
\label{fig:C2H4_QFSM} 
\end{figure*}

We also calculated the PECs for the four low-lying electronic states of ethylene along the internal rotation around the double bond. The cc-pVDZ basis set was used with $2$ electrons in $2$ active $\pi$ orbitals. As in the case of hydrogen, the convergence strongly depends on the initial guess. The following estimate states were used: 
\begin{equation}
\label{eq:state_QFMS_C2H4}
\begin{aligned}
    &|\Psi_{A_g} \rangle = |0011\rangle, \\
    &|\Psi_{B_{1u}} \rangle = |1001\rangle ,
\end{aligned}
\end{equation}
where orbital ordering is $|\pi_\beta^*\pi_\alpha^*\pi_\beta \pi_\alpha \rangle$. As before, CASCI$(2,2)$ energies were used as the energy shifts. The results are presented in Fig.~\ref{fig:C2H4_QFSM}.

At $0^\circ$, all states, except for the triplet $^3B_{1u}$, reached chemical accuracy within $50$ iterations. The triplet state likely converges with additional iterations. The convergence was fastest for the singlet $^1A_g$ states. At $90^\circ$, none of the states reached chemical accuracy within $50$ iterations, although the convergence trends were still visible. Up to $60^\circ$, both singlet $^1A_g$ states achieved chemical accuracy.

A key advantage of our GQSP-based QFSM implementation over variational quantum eigensolver VQE-based folded spectrum approaches is that we do not need to explicitly construct or measure the squared Hamiltonian. For instance, Ref.~\cite{cadi2024} showed that even for H$_2$ in the STO-$3$G basis, constructing a squared Hamiltonian requires the measurement of $24$ Pauli strings, and for H$_2$O, over $111,000$ Pauli strings are required. Conversely, the proposed approach does not necessitate additional Pauli measurements and completely avoids variational optimization. The main limitation of our method is that convergence is generally slower and is highly dependent on the quality of both the initial guess wave function and the target energy. Nevertheless, QFSM is a promising tool for accessing excited states, especially when a good initial state is available.

\section{Conclusion} 
\label{sec:conclusion}

We developed a family of GQSP-based polynomial filters for quantum state preparation, covering ground and excited state tasks through the QPI, QPL, QII, and QFSM schemes. 
A query complexity analysis shows that QPI and QPL achieve ground state preparation with near-optimal $O(1/(\gamma \Delta)) \log (1/\epsilon)$ scaling without requiring prior energy estimates, while QII becomes the most efficient option once a coarse energy approximation is available. 
QFSM extends the framework to excited states and remains practical when an energy guess lies within the local gap.

The GQSP construction enables these filters to be implemented with predictable resource requirements, avoiding Trotterization and relying only on a single block encoding of the Hamiltonian and straightforward amplitude amplification.
Numerical proof-of-concept simulations confirm the theoretical trends: QPL achieves consistent gains over QPI, QII improves over earlier inverse iteration methods, such as I-Iter and Q-Inv, and QFSM reaches excited states without variational optimization. 

The main limitation is the low success probability inherent to non-unitary filters, which necessitates postselection or amplitude amplification and thus increases circuit depth. 
In our analysis, these resource costs are evaluated in a fault-tolerant quantum computing setting, where coherent polynomial transformations and amplitude amplification can be implemented reliably.
At finite logical accuracy, increasing the polynomial degree produces a trade-off between improved spectral filtering and accumulated implementation error. 
The resulting accuracy floor depends on the success probability and spectral conditioning; close to a target eigenstate, however, the energy is stationary with respect to state perturbations, so state-preparation errors enter the energy only at second order to leading order.

Overall, these results show that classical power method intuition transfers cleanly to the quantum setting through GQSP, yielding simple, modular, and competitive state preparation algorithms suitable for both ground and excited states.

However, many aspects can be improved before the practical implementation of these methods.
First, eliminating the need for variational optimization in QPL by deriving polynomial coefficients from perturbation theory can improve efficiency. 
The QII method could benefit from replacing the current polynomial approximation of the inverse function with one that has a larger radius of convergence, potentially enabling accurate computation of excited states. 
In addition, imaginary energy shifts can be beneficial, and, more generally, Green's function methods can be implemented using this technique. 
In the case of QFSM, it would be advantageous to identify a function that better distinguishes energy levels, because it is currently common for multiple eigenvalues in the folded spectrum to cluster near $1$, greatly worsening convergence. 
Moreover, it is in principle possible to extend the present GQSP-based methods to block variants aimed at preparing several excited states simultaneously. 
Within the Bloch effective Hamiltonian framework, the corresponding wave operator can be written as an function of the Hamiltonian or of the perturbation, allowing its polynomial approximation through GQSP~\cite{Tong2021}. 
The main challenges involve constructing the required model space projectors and approximating resolvent-like functions whose spectral structure demands high degree polynomials~\cite{mitarai2023perturbation, li2025quantumalgorithmlowenergy}, thereby increasing the circuit depth and sensitivity to noise.
Finally, the integration of GQSP-based methods with other quantum-classical hybrid approaches may open the door for tackling larger and more complex molecular systems on near-term fault-tolerant quantum devices.

\begin{acknowledgments}
This study was supported by the MEXT Quantum Leap Flagship Program (MEXTQLEAP) Grant No. JPMXS0120319794; JST COI-NEXT Program, Grant No. JPMJPF2014 and JST ASPIRE Program JPMJAP2319.
This research was partially supported by the JSPS Grants-in-Aid for Scientific Research (KAKENHI) Grant No. JP23H03819.
N.Y. was supported by JST Grant Number JPMJPF2221, JST CREST Grant Number JPMJCR23I4, IBM Quantum, JST ASPIRE Grant Number JPMJAP2316, JST ERATO Grant Number JPMJER2302, and the Institute of AI and Beyond at the University of Tokyo.

\end{acknowledgments}

\appendix

\section{LCU block encoding}
\label{appendix:block_encoding}
The GQSP algorithm requires the signal operator to be a controlled unitary. Hence, before applying the GQSP to a Hamiltonian $H$, we must embed $H$ into a larger unitary operator. A commonly used method for achieving this is the linear combination of unitaries (LCU) approach~\cite{LCU}. The first step is to write the Hamiltonian as a linear combination of Pauli strings, for instance, by applying the Jordan--Wigner transformation to a second-quantized Hamiltonian:

\begin{equation}
\label{eq:lcu1}
H = \sum_{j=0}^{L-1} c_j P_j,
\end{equation}
where each $P_j$ is a tensor product of Pauli matrices (a Pauli string), and $c_j$ are the corresponding coefficients.
The value $L$ is the number of terms in the LCU decomposition.
 
The construction of LCU block encoding requires an auxiliary register of $\lceil \log_2(L) \rceil$ qubits to encode the coefficients $c_j$. Two key gate sequences are introduced:
\begin{itemize}
    \item \textsc{PREPARE}, which acts on the auxiliary register to create a superposition of $| j\rangle$ states weighted by $\sqrt{\tfrac{|c_j|}{\lambda_\text{LCU}}}$, 
    \item \textsc{SELECT}, which uses the $| j\rangle$ state of the auxiliary register to apply the corresponding Pauli operator $P_j$ with its phase on the main register.
\end{itemize}
Here, $\lambda_\text{LCU} = \sum_{j=0}^{L-1} |c_j|$ is the normalization factor that ensures the amplitudes $\sqrt{|c_j|/\lambda_\text{LCU}}$ form a valid quantum state in the \textsc{PREPARE} operation.
$|j\rangle$ denotes the computational basis state whose bit string encodes the integer $j$.

Concretely,
\begin{equation}
\label{eq:lcu2}
\text{PREPARE} |0\rangle = \sum_{j=0}^{L-1} \sqrt{\frac{|c_j|}{\lambda_\text{LCU}}} |j\rangle,
\end{equation}
and
\begin{equation}
\label{eq:lcu3}
\text{SELECT} = \sum_{j=0}^{L-1} |j\rangle \langle j| \otimes P_j.
\end{equation}

\begin{figure*}[ht]
  \centering
  \begin{adjustbox}{width=1\textwidth}
  \begin{quantikz}[column sep=0.5cm]
\lstick{$\ket{0}$}
   & \gate{R_Y(\theta_0)}\gategroup[2,steps=3,style={dashed,rounded
corners,fill=blue!20, inner
xsep=2pt},background,label style={label
position=above,anchor=north,yshift=+0.2cm}]{{\sc
PREPARE}} 
   & \octrl{1} & \ctrl{1} & \qw\gategroup[3,steps=7,style={dashed,rounded
corners,fill=red!20, inner
xsep=2pt},background,label style={label
position=above,anchor=north,yshift=+0.2cm}]{{\sc
SELECT}}  & \octrl{1} & \qw
   & \octrl{2} & \qw  & \octrl{2} & \qw 
   & \ctrl{1}\gategroup[2,steps=3,style={dashed,rounded
corners,fill=blue!20, inner
xsep=2pt},background,label style={label
position=above,anchor=north,yshift=+0.25cm}]{{\sc
PREPARE$^\dagger$}}   & \octrl{1} & \gate{R_Y^\dagger(\theta_0)} &\\
\lstick{$\ket{0}$}
   & \qw
   & \gate{R_Y(\theta_1)} & \gate{R_Y(\theta_2)} & \gate{X} & \gate{Z} & \gate{X}
   & \ctrl{1}  & \qw        & \octrl{1}      & \qw       
   & \gate{R_Y^\dagger(\theta_2)} & \gate{R_Y^\dagger(\theta_1)} & \qw      & \\
\lstick{$\ket{\psi}$}
   & \qw
   & \qw        & \qw        & \qw      & \qw        & \qw
   & \gate{X} & \gate{X}   & \gate{Z} & \gate{X}  
   & \qw        & \qw        & \qw     &  
\end{quantikz}
\end{adjustbox}
  \caption{Quantum circuit for an LCU block encoding of 
  $H = -a I + b X - c Z$. The dashed boxes indicate 
  \textsc{PREPARE} (blue) and \textsc{SELECT} (red).}
  \label{fig:LCU_circuit}
\end{figure*}

The full LCU operator is implemented as the sequence
$\text{PREPARE}^\dagger \;-\;\text{SELECT}\;-\;\text{PREPARE}$,
as illustrated in Fig.~\ref{fig:LCU_circuit}:
\begin{equation}
\begin{aligned}
U_{\mathrm{LCU}}
:=\;
(\text{PREPARE}^\dagger \otimes I)\,\text{SELECT} \\
\qquad\;\;\times(\text{PREPARE} \otimes I).
\end{aligned}
\end{equation}

Based on \cite{Low2019hamiltonian}, we can observe how the LCU operator acts on the Hamiltonian eigenvectors $| E_i\rangle$. By construction, it can be observed that $\operatorname{SELECT}^2=I$, therefore, for each $| 0\rangle | E_i\rangle$ state, there exists an orthogonal garbage state $| g_i\rangle$, such that:
\begin{equation}
\label{eq:lcu4}
\begin{aligned}
U_{\mathrm{LCU}} |0 \rangle |E_i\rangle
&= \frac{E_i}{\lambda_\text{LCU}}\, |0 \rangle |E_i\rangle
\\[-2pt]
&\quad
+ \sqrt{1 - \left( \frac{E_i}{\lambda_\text{LCU}} \right)^2}\, |g_i\rangle,
\\[6pt]
U_{\mathrm{LCU}} |g_i\rangle
&= \sqrt{1 - \left( \frac{E_i}{\lambda_\text{LCU}} \right)^2}\, |0 \rangle |E_i\rangle
\\[-2pt]
&\quad
- \frac{E_i}{\lambda_\text{LCU}}\, |g_i\rangle.
\end{aligned}
\end{equation}

This shows that $U_{\mathrm{LCU}}$ acts as a reflection in the two-dimensional invariant subspace spanned by $\{\lvert 0\rangle\lvert E_i\rangle,\,\lvert g_i\rangle\}_i$, with matrix:
\begin{equation}
\label{eq:lcu5}
U_{\mathrm{LCU}}
=
\begin{pmatrix}
\mathcal{H} & \sqrt{I-\mathcal{H}^{2}}\\[2pt]
\sqrt{I-\mathcal{H}^{2}} & -\mathcal{H}
\end{pmatrix},
\end{equation}
where $\mathcal{H} := H/\lambda_\text{LCU}$ is the normalized Hamiltonian (with eigenvalues $\tilde{E}_i = E_i/\lambda_\text{LCU}$).

For the purposes of QSP and GQSP, it is often more convenient to work with a rotation matrix rather than a reflection. This is achieved by multiplying $U_{\mathrm{LCU}}$ by a reflection around the $|0\rangle$ state on the ancilla register. Operationally, this acts trivially if the register is in $\lvert 0\rangle$ and flips the sign if the register is in any other state:
\begin{equation}
\label{eq:rotation_from_reflection}
\begin{aligned}
U_\text{Q} :=\;& ((2\ket{0}\!\bra{0} - I)\otimes I)\, U_{\mathrm{LCU}} \\[4pt]
=\;&
\begin{pmatrix}
\mathcal{H} & \sqrt{I-\mathcal{H}^{2}}\\[2pt]
-\sqrt{I-\mathcal{H}^{2}} & \mathcal{H}
\end{pmatrix}.
\end{aligned}
\end{equation}
This operator acts as an SU(2) rotation in each subspace, as required for Chebyshev-based polynomial transformations.

\section{Angle finding}
\label{appendix:angles}
A crucial component of any QSP-based algorithm is the determination of the rotation angles sequence. The GQSP algorithm is no exception; it uses a procedure very similar to that of conventional QSP and QSVT. An angle finding protocol typically involves four steps: \textit{truncation}, \textit{partition}, \textit{completion}, and \textit{carving}. Each step is described in detail below.

In GQSP, our goal is to apply an arbitrary function $f(z)$ to a unitary block encoding matrix $U$. As the first step, we approximate the target function using a complex-valued polynomial $F(z)$, for instance, using the Remez exchange algorithm. This procedure is called truncation: restricting the function to a finite-degree polynomial that approximates $f(z)$ sufficiently well for our purposes.

Once we obtain $F(z)$, the next step is partition, in which we rescale $F(z)$ to ensure that its absolute value does not exceed $1$ on a unit circle in the complex plane. Typically, this is achieved by dividing $F(z)$ by a sufficiently large constant $\alpha$, and defining $P(z) \;=\; F(z)/\alpha$. This results in $|P(z)| \leq 1$ on the complex unit circle.

Having obtained $P(z)$, we must then complete the polynomial to form the unitary that GQSP applies to $U$. Specifically, we seek a polynomial $Q(z)$ such that

\begin{equation}
    \label{eq:angle1}
    \operatorname{GQSP}(U) = 
    \begin{pmatrix}
        P(U)& * \\
        Q(U)& *
    \end{pmatrix}
\end{equation}
with the constraint $\lvert P(z)\rvert^2 + \lvert Q(z)\rvert^2 = 1$ on a unit circle. Several methods can be used to find $Q(z)$; four of the most common are via optimization, via root-finding, via Prony’s method, and the recently proposed exact expression using contour integration \cite{berntson2024complementarypolynomialsquantumsignal}.

The optimization-based technique used by the original GQSP authors~\cite{Motlagh2024} begins by applying a Fourier transform to the condition $\lvert P(z)\rvert^2 + \lvert Q(z)\rvert^2 = 1$. Denoting by $\vec{a}$ and $\vec{b}$ the coefficient vectors of $P(z)$ and $Q(z)$ respectively, one obtains
\begin{equation}
    \label{eq:angle2}
    \vec{a} \star \text{reverse}(\vec{a})^* + \vec{b} \star \text{reverse}(\vec{b})^* = \vec{\delta},
\end{equation}
where $\star$ denotes discrete convolution and $\text{reverse}(\cdot)$ refers to the reversal order of elements of an vector.

One can then solve
\begin{equation}
\label{eq:angle3}
\arg\min_{b}
\left\lVert
\vec{a} \star \mathrm{reverse}\bigl(\vec{a}\bigr)^{*} +
\vec{b} \star \mathrm{reverse}\bigl(\vec{b}\bigr)^{*} -
\vec{\delta}
\right\rVert^{2},
\end{equation}
determining a suitable $\vec{b}$. Although straightforward to implement, this method can be sensitive to initial guesses and may not be precise enough.

A more direct approach uses polynomial properties in a complex plane~\cite{chao2020findinganglesquantumsignal}. One first defines
\begin{equation}
    \label{eq:angle4}
    g(z) = 1 - |P(z)|^2. 
\end{equation}

Half of the roots of $g(z)$ lie inside the unit circle, whereas the other half lie outside. Let $\{\xi_j\}$ be the roots within the unit circle. Subsequently, a polynomial $Q(z)$ satisfying $\lvert Q(z)\rvert^2 = g(z)$ on $\lvert z\rvert=1$ can be written up to scaling as follows:
\begin{equation}
    \label{eq:angle5}
    Q(z) \propto \prod_{|\xi_j| < 1} (z - \xi_j). 
\end{equation}
Although conceptually simpler, this requires finding all roots of a potentially high-degree polynomial, making it computationally expensive.

Prony’s method was proposed to address the scaling challenges of root finding~\cite{Ying2022stablefactorization, yamamoto2024robustanglefindinggeneralized}. Define
\begin{equation}
    \label{eq:angle6}
    h(z) = \frac{1}{g(z)} = \frac{1}{1 - |P(z)|^2}.
\end{equation}
In this formulation, the roots of $g(z)$ become poles of $h(z)$. Hence $h(z)$ can be expressed as follows:
\begin{equation}
\label{eq:angle7}
    h(z) = \sum_{\xi_j} \frac{r_j}{z - \xi_j} + \text{constant},
\end{equation}
where $r_j$ are the residues of $h(z)$ at the poles $\xi_j$.
Next, one computes the Fourier coefficients $\hat{h}_k$ of $h(z)$:
\begin{equation}
\label{eq:angle8}
   \hat{h}_k = \frac{1}{2 \pi i} \int_\mathrm{G} \frac{h(z)}{z^k} \frac{dz}{z} = -\sum_{|\xi_j|<1} r_j \xi_j^{-(k+1)},
\end{equation}
where $\mathrm{G}$ denotes a unit circle contour. Thus, each $\hat{h}_k$ is a linear combination of the inverses of the roots inside $\mathrm{G}$. 
Following Prony’s approach, the polynomial for $Q(z)$ is obtained by solving
\begin{equation}
\label{eq:angle9}
H_{\text{Prony}}
\begin{pmatrix}
m_0 \\
m_1 \\
\vdots \\
m_d
\end{pmatrix}
= 0,
\end{equation}
where $H_{\text{Prony}}$ is the $l \times (d+1)$ Hankel matrix
\begin{equation}
\label{eq:angle95}
H_{\text{Prony}} =
\begin{pmatrix}
\hat{h}_{-1} & \hat{h}_{-2} & \cdots & \hat{h}_{-(d+1)} \\
\hat{h}_{-2} & \hat{h}_{-3} & \cdots & \hat{h}_{-(d+2)} \\
\vdots       & \vdots       & \ddots & \vdots           \\
\hat{h}_{-l} & \hat{h}_{-(l+1)} & \cdots & \hat{h}_{-(d+l)}
\end{pmatrix}.
\end{equation}
The vector $\vec{m} = (m_0, m_1, \dots, m_d)^{T}$ contains the coefficients of the $Q(z)$ up to a scaling factor and can be obtained using the singular value decomposition (SVD). In particular, $H_{\text{Prony}} = U_{\text{SVD}} \Sigma V^\dagger_{\text{SVD}}$, where $U_{\text{SVD}}$ and $V_{\text{SVD}}$ are unitary matrices, and $\Sigma$ is a diagonal matrix of singular values of $H_{\text{Prony}}$. Because $\mathrm{rank}(H_{\text{Prony}}) < d + 1$, $\Sigma$ has at least one zero and the corresponding column of $V_{\text{SVD}}$ supplies $\vec{m}$.

The most advanced technique for the completion step, which relies on complex analysis, was introduced in \cite{berntson2024complementarypolynomialsquantumsignal}. In this approach, one begins with the function $g(z)$ defined in Eq.~\eqref{eq:angle4}; however, this time, the logarithm is applied:
\begin{equation}
\label{eq:angle13}
   w(z) = \log (g(z)).
\end{equation}

Next, the Fourier series of the boundary function $w(e^{i\theta})$ is calculated, denoted by $\hat{w}(\theta)$. Then the projection operator $\Pi$ is applied, which leaves only non-negative frequency components from $\hat{w}(\theta)$ and halves the constant term: 
\begin{equation}
\label{eq:angle14}
\Pi\bigl[e^{i n \theta}\bigr]
:= 
\begin{cases}
e^{i n \theta}, & n \in \mathbb{Z}_{>0},\\
\dfrac{1}{2},   & n = 0,\\
0,               & n \in \mathbb{Z}_{<0}.
\end{cases}
\end{equation}
This projection ensures that $\Pi w(z)$ is holomorphic inside the unit disc; thus, the resulting polynomial $Q(z)$ will contain only non-negative powers of $z$.

Finally, the polynomial $Q(z)$ is recovered by exponentiating the projected function:
\begin{equation}
\label{eq:angle15}
   Q(z) = \exp [\Pi w(z)].
\end{equation}

After completing $P(U)$ and $Q(U)$, the final step is called carving, in which the GQSP sequence (see Eq.~\eqref{eq:gqsp3}) was unwrapped from the end:
\begin{equation}
\label{eq:angle11}
\begin{aligned}
    &{\begin{pmatrix}
        \hat{P}(U)& \cdot \\
        \hat{Q}(U)& \cdot
    \end{pmatrix}}
    =
    \operatorname{CU}_0^\dagger R(\theta_j,\phi_j,0)^\dagger
    {\begin{pmatrix}
        P(U)& \cdot \\
        Q(U)& \cdot
    \end{pmatrix}}
    \\
    = 
    &{\begin{pmatrix}
        e^{-i \phi_j} \cos (\theta_j) U^\dagger& \sin(\theta_j) U^\dagger \\
        e^{-i \phi_j} \sin (\theta_j) I& -\cos(\theta_j) I
    \end{pmatrix}}
    {\begin{pmatrix}
        P(U)& \cdot \\
        Q(U)& \cdot
    \end{pmatrix}}
    ,
\end{aligned}
\end{equation}
yielding new polynomials $\hat{P}(U)$ and $\hat{Q}(U)$ of one degree lower than $P(U)$ and $Q(U)$. At each iteration, the rotation angles are extracted using
\begin{equation}
\label{eq:angle12}
    \begin{aligned}
        &\theta_j = \tan^{-1}\bigg(\frac{|b_j|}{|a_j|} \bigg), \\
        &\phi_j = \arg \bigg( \frac{a_j}{b_j} \bigg), \\
        &\lambda_0 = \arg(b_0).
    \end{aligned}
\end{equation}
This process is repeated until the sequence is fully carved out, and any remaining global phase is stored in the final angle $\lambda_0$.

In practice, numerical problems may arise when $P(z)$ has nearly zero coefficients at high or low powers. This reduces the numerical rank of $H_{\text{Prony}}$ in Eq.~\eqref{eq:angle9}. To mitigate this, a \textit{capitalization} technique is used, in which an auxiliary polynomial $P'(z)$ with large coefficients at the highest and lowest powers is added and then subtracted:
\begin{equation}
\label{eq:angle10}
    P''(z) = \frac{1}{\beta} P(z) - P'(z) = \frac{1}{\beta} P(z) - \Upsilon (z^d + 1),
\end{equation} 
where $\beta$ and $\Upsilon$ are suitably chosen large constants. Subsequently, we work with $P''(z)$, which is numerically more stable.

Once $P''(U)$ is computed, the $P(U)$ must be recovered. This requires the addition of $P'(U)$ using an extra qubit, as illustrated in Fig.~\ref{fig:capitalization_circuit}:

\begin{figure}[ht]
  \centering
  \begin{quantikz}[row sep=0.1cm]
    \lstick{$\ket{0}$}
    & \gate{\mathrm{Had}}
    & \octrl{1}
    & \ctrl{1}
    & \gate{\mathrm{Had}} 
    & \ket{0} \\
    \lstick{}
    & 
    & \gate[3]{\begin{tabular}{c}
        GQSP \\
        for \\
        $P'(U)$
    \end{tabular}}
    & \gate[3]{\begin{tabular}{c}
        GQSP \\
        for \\
        $P''(U)$
    \end{tabular}}
    &
    & \\
    \lstick{}
    \quad \ldots \\
    \lstick{}
    & 
    &
    &
    &
    &
  \end{quantikz}
  \caption{Quantum circuit to restore the original polynomial $P(U)$ after the capitalization step, using LCU with one additional qubit. Had denotes Hadamard gates.}
  \label{fig:capitalization_circuit}
\end{figure}

Although this method doubled the circuit depth and consumed an additional qubit, it stabilized the numerical procedure.

\section{Comparison of QII with Q-Inv and I-Iter for different truncation}
\label{appendix:qii_truncation}

\begin{figure*}[ht] 
\includegraphics[width=1\textwidth]{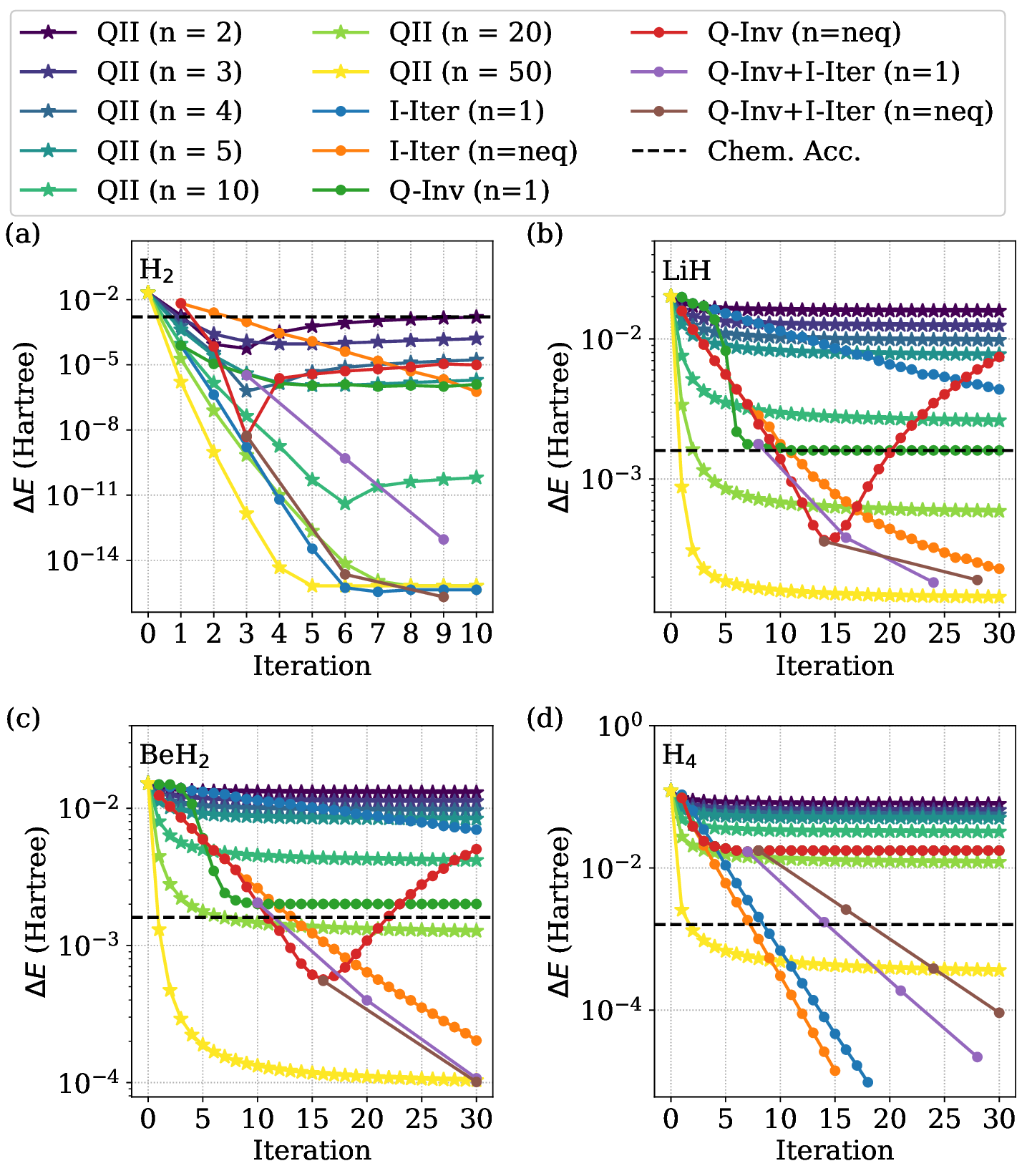} \caption{Comparison of QII, I-Iter, and Q-Inv \cite{Cainelli2024} for H$_2$, LiH, BeH$_2$, and H$_4$. Shown is the deviation from CASCI energies as a function of the iteration number. For QII, $n$ refers to the polynomial truncation degree, while for I-Iter and Q-Inv, $n$ is the Gauss-Legendre polynomial order. $n_{\mathrm{eq}}$ is the order at which energy convergence is reached.} 
\label{fig:Comparison_QII} 
\end{figure*}

Here, we analyze how the convergence behavior of QII changes with different truncations of the Taylor polynomial (see Fig.~\ref{fig:Comparison_QII}). As a general rule, the convergence limit depends on the truncation: the higher the polynomial degree, the lower the energy. 

For H$_2$, QII achieved chemical accuracy for all tested values $n$. Q-Inv($n=1$) is very close to QII($n=5$). The curves for QII($n=20$), QII($n=50$), I-Iter($n=1$) and Q-Inv+I-Iter($n=n_{\mathrm{eq}}$) are grouped near $4 \times 10^{-16}$, which is the most likely consequence of machine precision. 

For LiH, polynomial truncation in QII becomes a more significant constraint. Chemical accuracy was achieved only for $n \ge 20$, whereas lower $n$ values converged to energies above the chemical accuracy. However, once $n$ is sufficiently large, QII converges faster than Q-Inv and I-Iter; $\mathrm{QII}(n=20)$ requires only two iterations to reach chemical accuracy, whereas $\mathrm{QII}(n=50)$ does so in a single iteration.

A similar trend is observed for BeH$_2$. QII with $n \ge 20$ achieved chemical accuracy with seven iterations at $n=20$ and in only one iteration at $n=50$. Again, $\mathrm{QII}(n=50)$ initially converges faster than I-Iter or Q-Inv. However, after approximately 30 iterations, the combined Q-Inv+I-Iter approach slightly exceeded $\mathrm{QII}(n=50)$. Extending QII to $n>50$ further improves the final accuracy.

The ground state of the square H$_4$ system exhibits a strong multireference character, making it the most challenging case. Only $\mathrm{QII}(n=50)$ achieved chemical accuracy in only two iterations. However, at longer iteration counts (beyond $10$), I-Iter eventually reached lower energies. Similarly, increasing the degree of polynomial $n$ in QII can further enhance its performance. Conversely, the Q-Inv method alone did not achieve chemical accuracy of this system within the tested iterations.

\section{Pseudocodes}
\label{appendix:pseudo}

\begin{algorithm}[h]
\caption{Quantum power Lanczos (QPL$k$)}
\label{alg:QPL}
\begin{algorithmic}[1]
\REQUIRE Hamiltonian $\mathcal{H}$, initial state $\ket{\psi^{(0)}}$, Krylov order $d$, iterations $N$
\STATE Initialize coefficients $C_i$.
\STATE Define the Krylov polynomial
\[
    P_d(x) = \sum_{i=0}^{d} C_i x^i .
\]
\STATE \textbf{Variational optimization:}
\REPEAT
    \STATE Use GQSP to prepare
    \[
        \ket{\Psi_k^{(0)}} \propto P_d(\mathcal{H}) \ket{\psi^{(0)}} .
    \]
    \STATE Estimate the energy
    \[
        E^{(0)}_k = \bra{\Psi_k^{(0)}} \mathcal{H} \ket{\Psi_k^{(0)}} .
    \]
    \STATE Update $\{C_i\}$ to minimize $E^{(0)}_k$.
\UNTIL{convergence of $E^{(0)}_k$}
\STATE \textbf{Power iteration:}
\STATE $n=1$
\REPEAT
\STATE Use GQSP to prepare
    \[
        \ket{\Psi_k^{(n)}} \propto \mathcal{H}^n P_d(\mathcal{H}) \ket{\psi^{(0)}} .
    \]
\STATE Apply AA circuit
\STATE Estimate the energy
    \[
        E^{(n)}_k = \bra{\Psi_k^{(n)}} \mathcal{H} \ket{\Psi_k^{(n)}} .
    \]
\STATE $n = n + 1$
\UNTIL{$n=N+1$ or convergence of $E^{(n)}_k$}
\STATE \textbf{return} $\ket{\Psi_k^{(n)}}$ (and $E^{(n)}_k$)
\end{algorithmic}
\end{algorithm}

\begin{algorithm}[h]
\caption{Quantum inverse iteration (QII)}
\label{alg:QII}
\begin{algorithmic}[1]
\REQUIRE Hamiltonian $\mathcal{H}$, shift $\epsilon_0$, truncation order $M$, iterations $N$
\STATE $n=1$
\REPEAT
\STATE Compute polynomial coefficients $\{a_k\}_{k=0}^{M}$ approximating
\[
    (\mathcal{H} - \epsilon_0 I)^{-n} \approx \sum_{k=0}^{M} a_k \mathcal{H}^k .
\]
\STATE Use GQSP to prepare
    \[
        \ket{\Psi^{(n)}} \propto \sum_{k=0}^{M} a_k \mathcal{H}^k \ket{\psi^{(0)}} .
    \]
\STATE Apply AA circuit
\STATE Estimate the energy
    \[
        E^{(n)} = \bra{\Psi^{(n)}} \mathcal{H} \ket{\Psi^{(n)}} .
    \]
\STATE $n = n + 1$
\UNTIL{$n=N+1$ or convergence of $E^{(n)}$}
\STATE \textbf{return} $\ket{\Psi^{(n)}}$ (and $E^{(n)}$)
\end{algorithmic}
\end{algorithm}

\begin{algorithm}[h]
\caption{Quantum folded spectrum method (QFSM)}
\label{alg:QFSM}
\begin{algorithmic}[1]
\REQUIRE Hamiltonian $\mathcal{H}$, target shift $\epsilon_m$, constant $C>0$, iterations $N$
\STATE $n=1$
\REPEAT
\STATE Define the shifted Hamiltonian
\[
    \mathcal{H}_{\text{shifted}} = \mathcal{H} - \epsilon_m I .
\]
\STATE Define the folded-spectrum polynomial
\[
    F_n(x) = (1 - C x^2)^n .
\]
\STATE Use GQSP to prepare
    \[
        \ket{\Psi^{(n)}} \propto F_n(\mathcal{H}) \ket{\psi^{(0)}} .
    \]
\STATE Apply AA circuit
\STATE Estimate the energy
    \[
        E^{(n)} = \bra{\Psi^{(n)}} \mathcal{H} \ket{\Psi^{(n)}} .
    \]
\STATE $n = n + 1$
\UNTIL{$n=N+1$ or convergence of $E^{(n)}$}
\STATE \textbf{return} $\ket{\Psi^{(n)}}$ (and $E^{(n)}$)
\end{algorithmic}
\end{algorithm}

\bibliography{main.bib}% Include main.bib file

\end{document}